# ON THE FUNDAMENTAL EQUATION OF NONEQUILIBRIUM STATISTICAL PHYSICS

## ----NONEQUILIBRIUM ENTROPY EVOLUTION EQUATION AND THE FORMULA FOR ENTROPY PRODUCTION RATE

XING XIU-SAN
( Department of Physics, Beijing Institute of Technology, Beijing 100081, China )
(email:xingxiusan@gmail.com)

**Abstract** In this paper we presented an overview on our works. More than ten years ago, we proposed a new fundamental equation of nonequilibrium statistical physics in place of the present Liouville equation. That is the stochastic velocity type's Langevin equation in 6N dimensional phase space or its equivalent Liouville diffusion equation. This equation is time-reversed asymmetrical. It shows that the form of motion of particles in statistical thermodynamic systems has the drift-diffusion duality, and the law of motion of statistical thermodynamics is expressed by a superposition of both the law of dynamics and stochastic velocity and possesses both determinism and probability. Hence it is different with the law of motion of particles in dynamical systems. Starting from this fundamental equation the BBGKY diffusion equation hierarchy, the Boltzmann collision diffusion equation, the hydrodynamic equations such as the mass drift-diffusion equation, the Navier-Stokes equation and the thermal conductivity equation have been derived and presented here. What is more important, we first constructed a nonlinear evolution equation of nonequilibrium entropy density in 6N, 6 and 3 dimensional phase space, predicted the existence of entropy diffusion. This evolution equation reveals that the time rate of change of nonequilibrium entropy density originates together from its drift, diffusion and production in space. Furthermore, we presented a formula for entropy production rate (i.e. the law of entropy increase), proved that internal attractive force in nonequilibrium system can result in entropy decrease while internal repulsive force leads to another entropy increase, obtained an unified theoretical expression for thermodynamic degradation and self-organizing evolution, and revealed that the entropy diffusion mechanism caused the system to approach to equilibrium.

Statistical physics contains two parts consisting of equilibrium state and nonequilibrium state. The concepts and methods for equilibrium statistical physics so far have matured and are almost perfect after more than one hundred years study. Nonequilibrium statistical physics, its goal is attempting to study and understand the evolution laws of macroscopic nonequilibrium systems in terms of the dynamics of a large number of microscopic particles, as an independent discipline has been paid great attention to widely only in recent forty and fifty years, now is still in the development stage.

All the macroscopic thermodynamic processes in nature are time-directional or irreversible, while the classical mechanics and quantum mechanics are reversible. During constructing a nonequilibrium statistical physics, the first challenge is facing the problem of the paradox of irreversibility[1-3]: why the microscopic dynamics is reversible while the macroscopic statistical thermodynamical process is irreversible? This paradox has puzzled many physicists since the time of Boltzmann. It is in the exist statistical physics as a specific showing: the time-reversed symmetrical Liouville equation has been long considered as the fundamental equation of statistical physics, and it is consistent with three equilibrium ensemble , i.e., microcanonical ensemble, canonical ensemble and the grand canonical ensemble, and it is also can be applied to calculate the entropy of the equilibrium system. However, when using it to deduce and explain the irreversibility of nonequilibrium macroscopic systems, the law of entropy increase and the time-reversed asymmetrical hydrodynamic equations etc, if not adding any assumptions, we can not obtain correct results, or even impossibly give correct results in a rigorous and unified fashion[1-9]. What on earth is the origin of the paradox of irreversibility? Is it because the law of statistical thermodynamics is essentially different from the law of dynamics? If yes, what is the difference between them? Whether the nonequilibrium statistical physics has a fundamental equation? If so, what is its form? Can it provide a unified framework of statistical physics including nonequilibrium state and equilibrium state? How can we derive rigorously the hydrodynamic equations from microscopic kinetics? Does nonequilibrium entropy obey any evolution equation? What is the form of this equation if it exists? What is the microscopic physical basis of entropy production rate、namely the law of entropy increase? Can it be described by a quantitative concise formula? Does the entropy of an isolated system always only increase and never decrease? Whether or not there is a power of entropy decrease in nature to resist the law of entropy increase? If yes, what is its microscopic dynamical mechanism? And what is its mathematical expression? Can be thermodynamic degradation and self-organizing evolution unified? How to unify? What mechanism is responsible for the processes of approach to equilibrium? How to quantitatively describe it? Can we answer all the above questions from a new fundamental equation in a unified way? In recent more than ten years the author[10-18] has had new exploration around these issues, and obtained all important results enumerated in abstract. This paper is an overview on our works. The block diagram in the last section of this paper is a diagrammatic summary of this overview.

## 1. New fundamental equation—the stochastic velocity type's Langevin equation in 6N dimensional phase space

Each main branch of theoretical physics has its own fundamental equations, such as the Newton equation of motion in classical mechanics, the Schrödinger equation in quantum mechanics, the Maxwell equations in electrodynamics and so on. They have two common features: one is their fundamentality, that is, they are a complete and concise mathematical description of basic laws and phenomena and were drawn from experiments and postulate. They can neither be derived from other fundamental equations, nor be explained distinctly why it is so. The other one is their leading role, that is, they are capable of formulating and deriving nearly all the related physical laws and secondary equations, widely explaining various phenomena, and even making some predictions without adding any extra basic assumption. The fundamental equations are the soul, core and framework of theoretical physics. We believe that as an important and independent main discipline of theoretical physics, it must exist also a fundamental equation in nonequilibrium statistical physics. What is the form of this fundamental equation is a core topic, which we must solve in the development of the rigorous and unified nonequilibrium statistical physics. As mentioned above, the Liouville equation is equivalent the Hamilton equation in 6N dimensional phase space and is time-reversed symmetrical. When it is applied to explain the basic physical properties of nonequilibrium statistical thermodynamics, more than one extra assumption are needed. In fact because all the macroscopic quantities are the average values of the microscopic quantities, without any extra assumptions it is impossible to derive any irreversible macroscopic evolution equations from the reversible microscopic Liouville equation. Compared with the Newton equation of motion, the Schrödinger equation and Maxwell equations, the Liouville equation as the fundamental equation of statistical physics is incomplete, no matter so far as their fundamentality or leading role. Hence the author believe that it is wiser to propose a new fundamental equation with consideration of the assumption starting from enriching fundamental laws rather than to add again some extra patched up assumptions later on the basis of the old fundamental equation. That is to say, we should postulate an equation reflecting the fundamental law of nonequilibrium statistical thermodynamics as the fundamental equation of nonequilibrium statistical physics. Whether or not this equation is correct, it will depend on that the experiments prove if it has the above mentioned fundamentality and leading role. What is the fundamental law of nonequilibrium statistical thermodynamics? That is: all real statistical thermodynamic processes in nature are time-directional or irreversible (called statistical thermodynamic time's direction or irreversibility for short). This in essence is the general formulation of the second law of thermodynamics. It is a fundamental law of the holistic evolution of a macroscopic system in nature. Hence it cannot be derived from or reduced to another fundamental law— the dynamical law, still less should be the approximate result of dynamics. The contradiction between dynamical reversibility and thermodynamic irreversibility just is the representation of difference in essence between the law of motion of individuals and the law of macroscopic holistic motion of a large populations of particles. Just according to the idea that the fundamental equation should reflect the fundamental law of nonequilibrium statistical thermodynamics—thermodynamic time’s direction, more than ten years ago, the author[10-14] proposed a new equation of time-reversed asymmetry in place of the present Liouville equation of time-reversed symmetry as the fundamental equation of nonequilibrium statistical physics. That is to say, we made an assumption: The law of motion of the particles in statistical thermodynamic system obeys the following stochastic velocity type’s Langevin equation in 6N dimensional phase space (Γ space )

$$\begin{cases} \dot{\boldsymbol{q}}_i = \nabla_{\boldsymbol{p}_i} H + \boldsymbol{\eta}_i(\boldsymbol{q}_i, t) \\ \dot{\boldsymbol{p}}_i = -\nabla_{\boldsymbol{q}_i} H \end{cases} \tag{1}$$

Where

$$\begin{cases} \langle \boldsymbol{\eta}_i(\boldsymbol{q}_i, t)\rangle = 0 \\ \langle \boldsymbol{\eta}_i(\boldsymbol{q}_i, t)\boldsymbol{\eta}_j(\boldsymbol{q}_j, t')\rangle = 2D_{\boldsymbol{q}_i\boldsymbol{q}_i}(\boldsymbol{q}_i)\delta_{\boldsymbol{q}_i\boldsymbol{q}_j}\delta(t-t') \end{cases} \tag{2}$$

$H = H(\boldsymbol{X}) = H(\boldsymbol{q},\boldsymbol{p}) = H(\boldsymbol{q}_1,\boldsymbol{q}_2,\ldots\boldsymbol{q}_N;\boldsymbol{p}_1,\boldsymbol{p}_2,\ldots\boldsymbol{p}_N)$ is Hamiltonian of the system, $\boldsymbol{X} = (\boldsymbol{q},\boldsymbol{p})$ is the state vector in 6N dimensional phase space, $\boldsymbol{q}$ and $\boldsymbol{p}$ are the set vectors $\boldsymbol{q} = (\boldsymbol{q}_1,\boldsymbol{q}_2,\cdots\boldsymbol{q}_N)$ and $\boldsymbol{p} = (\boldsymbol{p}_1,\boldsymbol{p}_2,\ldots\boldsymbol{p}_N)$; $\boldsymbol{q}_i$ and $\boldsymbol{p}_i$ are the generalized coordinates and momenta of the ith particles respectively. $D = \{D_{\boldsymbol{q}_i\boldsymbol{q}_j}\} = \left\{D_{\boldsymbol{q}_i\boldsymbol{q}_i}(\boldsymbol{q}_i)\delta_{\boldsymbol{q}_i\boldsymbol{q}_j}\right\} = \left\{D(\boldsymbol{q}_i)\delta_{\boldsymbol{q}_i\boldsymbol{q}_j}\right\}$ is the diffusion matrix in the coordinate subspace of phase space, $D_{q_iq_i} = D(\boldsymbol{q}_i) = D(\boldsymbol{q}_1) = D(\boldsymbol{q}_2) = \cdots = D(\boldsymbol{q}_N)$ is just the ordinary diffusion matrix of particle in three dimensional coordinates space, its matrix element $D_{ij} = D_{ji}$ is six and can be calculated within framework of the theory and measured from experiment. Equation (1) and (2) show that the generalized velocities of particles in the statistical thermodynamic systems are not deterministic again, an extra stochastic term $\boldsymbol{\eta}_i(\boldsymbol{q}_i, t)$ has been added even though the force acting on the particles is deterministic. In other words, that the equation describes the law of motion of the particles in statistical thermodynamic system is expressed by a superposition of both the Hamilton equation in $6N$ dimensional phase space and stochastic velocity. It possesses both determinism and probability. Hence it is different from the Hamilton equation. Why we specially call equation (1) the stochastic velocity type's Langevin equation? This is because it is different from the ordinary formulation[7,19], the stochastic term in the Langevin equation is stochastic velocity rather than stochastic force. We shall see in the following sections, if equation (1) is not the stochastic velocity type's Langevin equation, but the ordinary stochastic force type's Langevin equation, that is, if the stochastic velocity is replaced by the stochastic force in equation (1), then the operater $\nabla_q{}^2$ on the right-hand side of the Liouville diffusion equation (3) (6) should be replaced by $\nabla_p{}^2$. As a result, the operaters $\nabla_q$ and $\nabla_q{}^2$ in the terms including the diffusion coefficient $D$ on the right-hand side of all the following equations and formulas、especially equations (35) (44) (49)、formulas (55) (58) and equations (19) (21) (26) should be replaced by $\nabla_p$ and $\nabla_p{}^2$. This means that the entropy production、entropy diffusion、mass diffusion、thermal conductivity and viscosity force, all of them would occur in momentum space, but not coordinate space. Evidently, this does not agree with experiments.

Just the same as that the Liouville equation is equivalent to the Hamilton equation in 6N-dimensional phase space, it is easy to prove [20] that the temporal evolution equation of probability density being equivalent to the temporal evolution equation (1) of dynamical variable is as follows according to Fokker-Planck rule [10-14].

$$\frac{\partial \rho}{\partial t} = -\dot{\boldsymbol{X}} \cdot \nabla_{\boldsymbol{X}} \rho + \nabla_{\boldsymbol{q}} \cdot [\nabla_{\boldsymbol{q}} \cdot (D\rho) - \frac{1}{2}(\nabla_{\boldsymbol{q}} \cdot D)\rho] = -\nabla_{\boldsymbol{X}} \cdot (\dot{\boldsymbol{X}}_t \rho)$$
$$= [H, \rho] + \nabla_{\boldsymbol{q}} \cdot [\nabla_{\boldsymbol{q}} \cdot (D\rho) - \frac{1}{2}(\nabla_{\boldsymbol{q}} \cdot D)\rho] \tag{3}$$

Where

$$\dot{\boldsymbol{X}} = (\nabla_{\boldsymbol{p}} H, -\nabla_{\boldsymbol{q}} H), \quad -\nabla_{\boldsymbol{X}} \cdot (\dot{\boldsymbol{X}}\rho) = -\dot{\boldsymbol{X}} \cdot \nabla_{\boldsymbol{X}} \rho = [H, \rho] \tag{4}$$

$$\dot{\boldsymbol{X}}_t = \dot{\boldsymbol{X}} + \frac{1}{2}(\nabla_{\boldsymbol{q}} \cdot D) - \frac{\nabla_{\boldsymbol{q}} \cdot (D\rho)}{\rho} \tag{5}$$

or
$$\frac{\partial \rho}{\partial t} = [H, \rho] + D : \nabla_{\boldsymbol{q}} \nabla_{\boldsymbol{q}} \rho \qquad (\text{when } \nabla_{\boldsymbol{q}} \cdot D = 0 \ ) \tag{6}$$

$\rho = \rho(\boldsymbol{X}, t) = \rho(\boldsymbol{q}, \boldsymbol{p}, t) = \rho(\boldsymbol{q}_1, \boldsymbol{q}_2, \cdots, \boldsymbol{q}_N; \boldsymbol{p}_1, \boldsymbol{p}_2, \cdots, \boldsymbol{p}_N; t)$ is the ensemble probability density. Equation (3) may be called the Liouville diffusion equation. We assume exactly equation (1) or (3) to be the new fundamental equation of nonequilibrium statistical physics. Compared with the Liouville equation, this equation has an extra stochastic diffusive term in coordinate subspace, which represents that the particles in statistical thermodynamic system not only drifts in phase space but also diffuses in coordinate subspace. Hence it is microscopic time-reversed asymmetrical and reflects the irreversibility of statistical thermodynamic processes. The drift motion reflects reversible dynamic character, and the stochastic diffusion is microscopic origin of thermodynamic time's direction. That is to say, thermodynamic irreversibility is macroscopic representation of microscopic stochasticity of particles. The drift-diffusion duality of the particles means that the law of motion of statistical thermodynamics both is constrained by the law of dynamics and possesses the character of stochastic process simultaneously. Dynamics and stochasticity, both of them seem to be intrinsic and exist simultaneously, but never be reduced one to the other.

It should be emphasized here, introducing the velocity stochasticity into the fundamental law of statistical physics, and proposing the stochastic velocity type's Langevin Eq.(1) in 6N dimensional phase space or its equivalent Liouville diffusion Eq.(3) as the new fundamental equation of nonequilibrium statistical physics, this is only a fundamental postulate. In the following sections we shall see, just owing to this fundamental postulate, it both overcomes the incompleteness of the Liouville equation and avoids the shortcoming that starting from the Langevin equation[7,19] cannot derive directly the kinetic equations and hydrodynamic equations. We not only derived concisely the hydrodynamic equations. Furthermore, we first obtained nonequilibrium entropy evolution equation in 6N, 6 and 3 dimensional phase space, predicted the exsistence of entropy diffusion, presented a formula for entropy production rate and a formula for entropy decrease rate, revealed the entropy diffusion mechanism for approach to equilibrium, and unify thermodynamic degradation and self-organizing evolution. Up to now, we did not see any other theory in this discipline, which can obtain so many important results rigorously and in a unified way only from a fundamental postulate. As to why we use white noise, the reason is for simplicity of calculation.

By the way, from mathematical viewpoint the Liouville diffusion equation (3) (6) in fact is a special Fokker-Planck equation in 6N dimensional phase space, and the equations (12) is the Smoluchowski type's Fokker-Planck equation in 3 dimensional space when the interaction between two particles and the external force can be neglected.

It should be still pointed out here, to propose that the law of motion of the particles in statistical thermodynamic system obeys the stochastic velocity type's Langevin equation in **6N** dimensional phase space does not exclude that the law of motion of the ordinary Brownian particle obeys the stochastic force type's Langevin equation, because both of them have different physical meaning.

For the sake of affirming the practical significance of equation(1), let us consider a real model: a real walk model of drunken man. An drunken man walks along the city road. He from one intersection enters into some road and walks a distance to another intersection, then he enters into another road and walks another distance. He repeats this walk over and over again. Because through every one intersection there are many roads, the velocity that the drunken man enters into some road is stochastic. However, after entering into the road, the velocity that the drunken man walks on this distance is deterministic. Thus, a real walk model of drunken man reveals that the form of motion of the drunken man in space is equal to the algebraic sum of the deterministic velocity and the stochastic velocity. Equation (1) just describes this form of motion in space. When the deterministic velocity is zero, equation (1) reduces to the expression for the typical random walk model[1,2]. Here, the velocity stochasticity originates from that many roads go through one intersection, which makes that the space position of the drunken man in the process of motion suddenly stochastically changes in some point (intersection). We remark here again, for the ordinary random walk model, when it as a nonequilibrium statistical physical process is described by a dynamic variable equation, this equation just is the stochastic velocity type's Langevin equation (1) with zero deterministic part. Why the generalized velocity of particles in statistical thermodynamic systems has the similar stochastic term? There is no satisfactory explanation at present. In fact, a glass fell to many pieces, a heavy nucleus splits into two fragments, they all have the similar stochasticity. We neither know the individual origin of these stochasticities, nor know whether they have some common essential origin. However, we can affirm that all the irrevesibilities for the above statistical thermodynamic process, real walk process of drunken man, glass broken process and nuclear fission process are caused by the stochasticity. In other word, stochasticity is the common origin of the irrevesibilities for all above processes.

Corresponding to (6), the quantum Liouville diffusion equation is

$$\frac{\partial \rho}{\partial t} = H\rho - \rho H + D\nabla_{\boldsymbol{q}}^2 \rho = [H, \rho] + D\nabla_{\boldsymbol{q}}^2 \rho \tag{6q}$$

where $\rho$ is density operator.

According to the stochastic theory[21], we can define the total temporary derivative of the ensemble probability density

$$\begin{aligned} \frac{d\rho}{dt} &= \frac{\partial \rho}{\partial t} + \dot{\boldsymbol{X}}_t \cdot \nabla_{\boldsymbol{X}} \rho \\ &= \frac{\partial \rho}{\partial t} + (\dot{\boldsymbol{X}} + \frac{1}{2}\nabla_{\boldsymbol{q}} \cdot D) \cdot \nabla_{\boldsymbol{X}} \rho - \frac{1}{\rho}\left[(\nabla_{\boldsymbol{q}}\rho) \cdot \nabla_{\boldsymbol{q}} \cdot (D\rho)\right] \end{aligned} \tag{7}$$

Substituting Eq. (3) into Eq. (4), we have

$$\frac{d\rho}{dt} = \nabla_{\boldsymbol{q}} \nabla_{\boldsymbol{q}} : (D\rho) - \frac{(\nabla_{\boldsymbol{q}}\rho) \cdot [\nabla_{\boldsymbol{q}} \cdot (D\rho)]}{\rho} \tag{8}$$

It can be seen from Eq. (8) that $(d\rho/dt) \neq 0$ for nonequilibrium states, i.e. the ensemble probability density is not stationary along phase trajectories. It will be diffusive in the co-ordinate subspace of phase space till approaching to equilibrium state $(d\rho/dt) = 0$, which thoroughly reflects the irreversibility as an inherent nature of nonequilibrium processes.

## 2. The BBGKY diffusion equation hierarchy

As is well known, the BBGKY equation hierarchy[2,3] can be derived from the Liouville equation. Similarly, the BBGKY diffusion equation hierarchy[11,13] can also be derived from the Liouville diffusion equation. Practically, it can be done by transforming the diffusion term $D : \nabla_{\boldsymbol{q}} \nabla_{\boldsymbol{q}} \rho$ in Eq. (6) into the reduced term and adding it to the BBGKY equation hierarchy.

Hereafter we assume $\nabla_{\boldsymbol{q}} \cdot D = 0$, i.e. the diffusion matrix $D$ of the particle is independent on coordinate $\boldsymbol{q}$.

Introducing the reduced probability density for $S$ particles

$$f_s(\boldsymbol{x}_1, \boldsymbol{x}_2, \cdots \boldsymbol{x}_s, t) = \int \cdots \int \rho(\boldsymbol{X}, t) d\boldsymbol{x}_{s+1} \cdots d\boldsymbol{x}_N$$

$$\int f_s(\boldsymbol{x}_1, \boldsymbol{x}_2, \cdots \boldsymbol{x}_s, t) d\boldsymbol{x}_1 d\boldsymbol{x}_2 \cdots d\boldsymbol{x}_s = 1$$

where $\boldsymbol{x}_i = (\boldsymbol{q}_i, \boldsymbol{p}_i)$, $\boldsymbol{q}_i$ and $\boldsymbol{p}_i$ are the generalized coordinates and momenta of particle $i$. We can prove

$$\int (D : \nabla_{\boldsymbol{q}} \nabla_{\boldsymbol{q}}) \rho(\boldsymbol{X}, t) d\boldsymbol{x}_{s+1} \cdots d\boldsymbol{x}_N$$
$$= \sum_{i=1}^{S} (D : \nabla_{\boldsymbol{q}_i} \nabla_{\boldsymbol{q}_i}) f_s(\boldsymbol{x}_1, \boldsymbol{x}_2, \cdots \boldsymbol{x}_s, t) \tag{9}$$

Adding (9) to the BBGKY equation hierarchy yields the reduced equation hierarchy for the reduced probability density $f_s(\boldsymbol{x}_1, \boldsymbol{x}_2, \cdots \boldsymbol{x}_s, t)$ as follows

$$\frac{\partial f_s}{\partial t} + H_s f_s = (N - S) \sum_{i=1}^{S} \int (\nabla_{\boldsymbol{q}_i} \phi_{i,S+1}) \cdot [\nabla_{\boldsymbol{p}_i} f_{s+1}(\boldsymbol{x}_1, \boldsymbol{x}_2, \cdots \boldsymbol{x}_{s+1}, t)] d\boldsymbol{x}_{s+1}$$
$$+ \sum_{i=1}^{S} (D : \nabla_{\boldsymbol{q}_i} \nabla_{\boldsymbol{q}_i}) f_s(\boldsymbol{x}_1, \boldsymbol{x}_2, \cdots \boldsymbol{x}_s, t) \tag{10}$$

Where

$$H_s = -\sum_{i=1}^{S} [(-\boldsymbol{F}_i + \sum_{k=1}^{S} \nabla_{\boldsymbol{q}_i} \phi_{ik}) \cdot \nabla_{\boldsymbol{p}_i} - \frac{\boldsymbol{p}_i}{m} \cdot \nabla_{\boldsymbol{q}_i}] \tag{11}$$

and $\boldsymbol{F}_i$ is the external force on the system, $\phi_{ik} = \phi_{\boldsymbol{q}_i \boldsymbol{q}_k} = \phi(|\boldsymbol{q}_i - \boldsymbol{q}_k|)$ is a two-body interaction potential.

The most useful equations in the equation hierarchy (10) are those for the reduced probability density of the single particle $f_1(\boldsymbol{x}, t) = f_1(\boldsymbol{q}, \boldsymbol{p}, t)$ and the two particles $f_2(\boldsymbol{x}_1, \boldsymbol{x}_2, t) = f_2(\boldsymbol{q}_1, \boldsymbol{p}_1; \boldsymbol{q}_2, \boldsymbol{p}_2; t)$

$$[\frac{\partial}{\partial t} + \frac{\boldsymbol{p}}{m} \cdot \nabla_{\boldsymbol{q}} + \boldsymbol{F} \cdot \nabla_{\boldsymbol{p}}] f_1(\boldsymbol{x}, t) = N \int (\nabla_{\boldsymbol{q}} \phi_{\boldsymbol{q}\boldsymbol{q}_1}) \cdot \nabla_{\boldsymbol{p}} f_2(\boldsymbol{x}, \boldsymbol{x}_1, t) d\boldsymbol{x}_1$$
$$+ (D : \nabla_{\boldsymbol{q}} \nabla_{\boldsymbol{q}}) f_1(\boldsymbol{x}, t) \tag{12}$$

$$[\frac{\partial}{\partial t} + \frac{\boldsymbol{p}_1}{m} \cdot \nabla_{\boldsymbol{q}_1} + \frac{\boldsymbol{p}_2}{m} \cdot \nabla_{\boldsymbol{q}_2} + (\boldsymbol{F}_1 - \nabla_{\boldsymbol{q}_1} \phi_{\boldsymbol{q}_1 \boldsymbol{q}_2}) \cdot \nabla_{\boldsymbol{p}_1}$$
$$+ (\boldsymbol{F}_2 - \nabla_{\boldsymbol{q}_2} \phi_{\boldsymbol{q}_1 \boldsymbol{q}_2}) \cdot \nabla_{\boldsymbol{p}_2}] f_2(\boldsymbol{x}_1, \boldsymbol{x}_2, t)$$
$$= N \int [(\nabla_{\boldsymbol{q}_1} \phi_{\boldsymbol{q}_1 \boldsymbol{q}_3}) \cdot \nabla_{\boldsymbol{p}_1} + (\nabla_{\boldsymbol{q}_2} \phi_{\boldsymbol{q}_2 \boldsymbol{q}_3}) \cdot \nabla_{\boldsymbol{p}_2}] f_3(\boldsymbol{x}_1, \boldsymbol{x}_2, \boldsymbol{x}_3, t) d\boldsymbol{x}_3$$
$$+ D : (\nabla_{\boldsymbol{q}_1} \nabla_{\boldsymbol{q}_1} + \nabla_{\boldsymbol{q}_2} \nabla_{\boldsymbol{q}_2}) f_2(\boldsymbol{x}_1, \boldsymbol{x}_2, t) \tag{13}$$

where $D = D_{\boldsymbol{q}\boldsymbol{q}}$ is the diffusion tensor in three dimensional space.

Compared with the BBGKY equation hierarchy, the reduced equation hierarchy (10) (12) and (13)are added the diffusion terms. Hence, they are time-reversal

asymmetrical.They may be called the BBGKY diffusion equation hierarchy. Equation (12) may be reduced to the kinetic equation.

For dilute gases, by the assumption of molecular chaos[2,3]

$$f_2(\boldsymbol{x}_1,\boldsymbol{x}_2,t) = f_1(\boldsymbol{x}_1,t)f_1(\boldsymbol{x}_2,t) \tag{14}$$

then Eq.(12) becomes

$$\begin{aligned}[\frac{\partial}{\partial t} + \frac{\boldsymbol{p}}{m}\cdot\nabla_{\boldsymbol{q}} + \boldsymbol{F}\cdot\nabla_{\boldsymbol{p}}]f(\boldsymbol{q},\boldsymbol{p},t) &= D\nabla_{\boldsymbol{q}}^2 f(\boldsymbol{q},\boldsymbol{p},t) \\ &+\int \mathrm{g}\sigma(\mathrm{g},\theta)[f(\boldsymbol{q},\boldsymbol{p}',t)f(\boldsymbol{q},\boldsymbol{p}_1',t) \\ &-f(\boldsymbol{q},\boldsymbol{p},t)f(\boldsymbol{q},\boldsymbol{p}_1,t)]d\boldsymbol{p}_1 d\Omega \end{aligned} \tag{15}$$

Because that the left-hand side and the first term on the right-hand side of Eq.(12) becomes the Boltzmann equation[2,3]. Equation (15) may be called the Boltzmann collision diffusion equation. Its collision occurs in the momentum space and diffusion happens in the coordinate space, so it describes the evolution of particle distribution both in momentum space and in coordinate space(hereafter assume diffusion coefficient $D$ is a scalar).

If the gas is assumed to be in a spatially homogeneous state, that is, if we can neglect the coordinate distribution of the particles, $f(\boldsymbol{q},\boldsymbol{p},t) = f(\boldsymbol{p},t)$ is independent on $\boldsymbol{q}$, then $D\nabla_q^2 f(\boldsymbol{p},t) = 0$ and Eq.(15) reduces to the Boltzmann equation. In this case from its equilibrium solution we can obtained the Maxwell distribution.

It should be pointed out here that Eq.(15) do not have conventional collisional invariants, because the law of motion described by them possesses stochasticity. When any particle makes a drift collision with some other one, it also interacts by way of diffusion collision with the global system of particles at the same time. The latter is a dissipative processes which will dissipate momentum and energy. Hence the conventional binary collisional invariance of these equations are broken. However, the total (including drift and diffusion) mass, momentum and energy conservation laws are still valid.

If the diffusion term is neglected from the first for dilute gases, then Equation (15) should be the Boltzmann (collision) equation but not the Boltzmann collision diffusion equation. Thus the above discussion is superfluous.

## 3. Hydrodynamic equations

How can man derive rigorously the hydrodynamic equations from microscopic kinetics? This is an open important problem up to now. Although the Navier-Stokes equation can be obtained approximately from the Boltzmann equation, the fluids is not all the dilute gases. Hence the Boltzmann equation is not valid for them.

Now let us derive the hydrodynamic equations from the BBGKY diffusion Eqs.(12) and (13) succinctly as follows.

It is well known that the mass balance equation of fluids derived from the BBGKY equation hierarchy is[2,3]

$$\frac{\partial\rho}{\partial t} + \nabla\cdot(\rho\boldsymbol{C}) = 0 \tag{16}$$

Where $\nabla = \nabla_q$, $\rho = \rho(\boldsymbol{q},t)$ is mass density of fluids, $\boldsymbol{C} = \boldsymbol{C}(\boldsymbol{q},t)$ is the mean velocity of fluid. The momentum balance equation of fluids is derived as[2,3]

$$\frac{\partial(\rho \boldsymbol{C})}{\partial t} + \nabla\cdot(\rho \boldsymbol{CC} + P) = \rho \boldsymbol{F} \tag{17}$$

Where $P$ is the pressure tensor. The internal energy balance equation of fluid is derived as[3,4]

$$\frac{\partial(\rho u)}{\partial t} + \nabla\cdot(\rho u\boldsymbol{C} + \boldsymbol{J}_{\boldsymbol{q}}) = -P:\nabla\boldsymbol{C} \tag{18}$$

where $u = u(\boldsymbol{q}, t)$ is the internal energy density of fluids, $\boldsymbol{J}_{\boldsymbol{q}}$ is heat flow.

In order to obtain the hydrodynamic equations we only need to reduce the diffusion term on the right-hand side of Eqs.(12) and (13)to the fluid term and add it to the corresponding Eqs.(16)—(18) respectively, then obtain[11,13].

3.1 Fluid mass evolution equation

$$\frac{\partial \rho}{\partial t} + \nabla\cdot(\rho \boldsymbol{C}) = D\nabla^2 \rho \tag{19}$$

which describes that the time rate of change in mass density of fluids originates from both drift and diffusion flow. This equation may be also called the mass drift diffusion equation . It can be represented as mass density continuity equation

$$\frac{\partial \rho}{\partial t} = -\nabla\cdot\boldsymbol{j} = -\nabla\cdot(\rho\boldsymbol{C} - D\nabla\rho) \tag{19a}$$

Where $\boldsymbol{j}$ is the mass density current.

When the diffusion term can be neglected, Eq.(19) reduces to Eq.(16). When the drift term can be neglected, Eq.(19) reduces to the ordinary diffusion equation:

$$\frac{\partial \rho}{\partial t} = D\nabla^2 \rho \tag{20}$$

3.2 Fluid momentum evolution equation

$$\frac{\partial(\rho \boldsymbol{C})}{\partial t} + \nabla\cdot(\rho \boldsymbol{CC} + P) = \rho \boldsymbol{F} + D\nabla^2(\rho\boldsymbol{C}) \tag{21}$$

Subtracting (19) multiplied both sides by $\boldsymbol{C}$ from Eq.(21) we obtain

$$\rho\frac{\partial \boldsymbol{C}}{\partial t} + \rho(\boldsymbol{C}\cdot\nabla)\boldsymbol{C} + \nabla\cdot P = \rho\boldsymbol{F} + \eta\nabla^2\boldsymbol{C} + \rho(2\nu\nabla\ln\rho)\cdot\nabla\boldsymbol{C} \tag{22}$$

or

$$\frac{\partial \boldsymbol{C}}{\partial t} + (\boldsymbol{C}\cdot\nabla)\boldsymbol{C} + \frac{1}{\rho}\nabla\cdot P == \nu\nabla^2\boldsymbol{C} + [(2\nu\nabla\ln\rho)\cdot\nabla]\boldsymbol{C} \tag{23}$$

where the viscosity coefficient

$$\eta = \rho D \tag{24}$$

and the kinematic viscosity coefficient

$$\nu = D = \eta / \rho \tag{25}$$

Equation(23) is the generalized Navier-Stokes equation ($\boldsymbol{F} = 0$).It is different from the Navier-Stokes equation due to having the additional term $[(2\nu\nabla\ln\rho)\cdot\nabla]\boldsymbol{C}$, which is caused by the mass density gradient $\nabla\rho$ and which is perhaps valuable .When $\nabla\rho$ is zero , Eq.(23) reduces to the Navier-Stokes equation .The formula (24) gives the relation between the viscosity coefficient $\eta$ and the diffusion coefficient $D$.The formula (25) shows that the kinematic viscosity coefficient $\nu$ is just the diffusion coefficient $D$,both of them are equal in magnitude and different in physical meaning .

3.3 Fluid internal energy evolution equation

$$\frac{\partial(\rho u)}{\partial t}+\nabla\cdot(\rho u\boldsymbol{C}+\boldsymbol{J}_q)$$
$$=-P:\nabla\boldsymbol{C}+D\nabla^2(\rho u)+\rho D(\nabla\boldsymbol{C}):(\nabla\boldsymbol{C})^T \qquad (26)$$

Subtracting (19) multiplied both side by $u$ from (26) we obtain

$$\rho\frac{\partial u}{\partial t}+\rho(\boldsymbol{C}\cdot\nabla)u+\nabla\cdot\boldsymbol{J_q}$$
$$=-P:\nabla\boldsymbol{C}+\rho D\nabla^2 u+\rho D(\nabla\boldsymbol{C}):(\nabla\boldsymbol{C})^T+(2D\nabla\rho)\cdot\nabla u \qquad (27)$$

or

$$\frac{\partial u}{\partial t}+(\boldsymbol{C}\cdot\nabla)u+\frac{1}{\rho}\nabla\cdot\boldsymbol{J}_q=-\frac{1}{\rho}P:\nabla\boldsymbol{C}+D\nabla^2 u$$
$$+D(\nabla\boldsymbol{C}):(\nabla\boldsymbol{C})^T+(2D\nabla\ln\rho)\cdot(\nabla u) \qquad (28)$$

Introducing the local temperature $T=T(\boldsymbol{q},t)$ and substituting the transformations ($\partial u/\partial t=C_V(\partial T/\partial t)$) and $\nabla u=C_V\nabla T$ into Eq. (28), we obtain the local temperature evolution equation of fluids

$$\frac{\partial T}{\partial t}+(\boldsymbol{C}\cdot\nabla)T+\frac{1}{\rho C_V}\nabla\cdot\boldsymbol{J}_q=\frac{\lambda}{\rho C_V}\nabla^2 T+\frac{D}{C_V}(\nabla\boldsymbol{C}):(\nabla\boldsymbol{C})^T$$
$$+(2D\nabla\ln\rho)\cdot(\nabla T) \qquad (29)$$

where $C_V$ is the specific heat per unit mass at constant volume and

$$\lambda=\rho C_V D \qquad (30)$$

is the thermal conductivity tensor.

Comparing Eqs. (19), (21) and (26) with Eqs. (16)—(18), we can see that there are not only the mass drift term $\nabla\cdot(\rho\boldsymbol{C})$, the momentum drift term $\nabla\cdot(\rho\boldsymbol{CC})$ and the internal energy drift term $\nabla\cdot(\rho u\boldsymbol{C})$, but also the mass diffusion term $D\nabla^2\rho$, the momentum diffusion $D\nabla^2(\rho\boldsymbol{C})$ and the internal energy diffusion term $D\nabla^2(\rho u)+\rho D(\nabla\boldsymbol{C}):(\nabla\boldsymbol{C})^T$ in the hydrodynamic equations derived in this paper. It may be seen from (21) and (23) that the emergence of the fluid viscosity is result of the momentum diffusion and hence the generalized Navier-Stokes equation is derived succinctly and rigorously. The mass diffusion, viscosity and heat conductivity are all the irreversible dissipative processes which are related to the stochasticity closely. The time-reversal asymmetry of hydrodynamic Eqs. (19), (23), (28) and (29) just represent irreversibility of these processes.

## 4. Nonequilibrium entropy evolution equation

Entropy is one of the most important concept and physical quantity in physics. The entropy change represents the evolution direction of macroscopic nonequilibrium physical systems. Although the entropy and the law of entropy increase have been extensively studied[5,22-28], a lot of major physical properties of nonequilibrium entropy have not been fully understood so far. From the viewpoint of microscopic statistical theory, in all the field of entropy, since for a long time past, we only have a Boltzmann's entropy formula which gives microscopic statistical explanation of entropy, but did not have any statistical equation and formula to describe the change of nonequilibrium entropy. This just remains to be explored and solved. As mentioned above, the mass, energy and momentum in nonequilibrium statistical thermodynamic system not only have their balance equations, but also all obey a evolution equation changing with time and space, such as diffusion equation, thermal conduction equation and the Navier-Stokes equation. Nonequilibriurn entropy, it has balance equation long ago.[3,27] The question naturally arises: How it changes with time and space? Does it also obey any evolution equation? What is the form of this equation if it exist? More than ten years ago, the author[12-16] first wrote down a nonlinear evolution equation of nonequilibriurn entropy density changing with time and space in 6N, 6 and 3 dimensional phase space, and predicted the existence of entropy diffusion .The results are given as follows.

4.1 6N dimensional nonequilibriurn entropy evolution equation

The nonequilibrium entropy in 6N-dimensional phase space can be defined as[7,11-16,27]

$$S_G(t) = -k\int \rho(\boldsymbol{X},t)\ln\frac{\rho(\boldsymbol{X},t)}{\rho_0(\boldsymbol{X})}d\Gamma + S_{G0}$$

$$= \int S_X d\Gamma + S_{G0} \tag{31}$$

where $k$ is the Boltzmann constant, $\rho_0$ and $S_{G0}$ are the ensemble probability density and the entropy for the equilibrium states respectively, $\rho_0$ satisfies

$$\frac{\partial\rho_0}{\partial t} = [H,\rho_0] + D\nabla_{\boldsymbol{q}}^2\rho_0 = 0 \tag{6a}$$

The entropy density in 6N dimensional phase space

$$S_X = -k\rho\ln\frac{\rho}{\rho_0} \tag{32}$$

or

$$S_G(t) = -k\int \rho(\boldsymbol{X},t)\ln\rho(\boldsymbol{X},t)d\Gamma = \int S_X d\Gamma \tag{31a}$$

$$S_X = -k\rho(\boldsymbol{X},t)\ln\rho(\boldsymbol{X},t) \tag{32a}$$

In this paper, we use formulas (31),(40) rather than (31a),(40a) as the definition of nonequilibrium entropy, the reason will be explained by the following formulas (58a)(96a) in section 6.

Differentiating both sides of formula (31) with respect to time $t$ and using the Liouville diffusion Eqs. (6) and (6a), we obtain the time rate of change of 6N dimensional nonequilibrium entropy

$$\frac{\partial S_G}{\partial t} = \int\frac{\partial S_X}{\partial t}d\Gamma = -k\int\left\{[\frac{\partial\rho}{\partial t}(\ln\frac{\rho}{\rho_0}+1)] - \frac{\rho}{\rho_0}\frac{\partial\rho_0}{\partial t}\right\}d\Gamma$$

$$= -\int[\nabla_X\cdot(\boldsymbol{J}_{st}+\boldsymbol{J}_{sd}) - \sigma_G]d\Gamma \tag{33}$$

and the balance equation of the entropy density in 6N-dimensional phase space

$$\frac{\partial S_X}{\partial t} = -\nabla_X \cdot \boldsymbol{J}_s + \sigma_G = -\nabla_X \cdot (\boldsymbol{J}_{st} + \boldsymbol{J}_{sd}) + \sigma_G \tag{34}$$

Hence we find out the evolution equation of nonequilibrium entropy density in 6N dimensional phase space. [12-16]

$$\frac{\partial S_X}{\partial t} = -\nabla_X \cdot (\dot{\boldsymbol{X}} S_X) + D\nabla_{\boldsymbol{q}}^2 S_X + \frac{D}{k\rho}[(\nabla_{\boldsymbol{q}} \ln \rho) S_X - \nabla_{\boldsymbol{q}} S_X]^2 \tag{35}$$

Where $-\nabla_X \cdot (\dot{\boldsymbol{X}} S_X) = [H, S_X]$

The entropy flow density

$$\boldsymbol{J}_s = \boldsymbol{J}_{st} + \boldsymbol{J}_{sd} = \dot{\boldsymbol{X}} S_X - D\nabla_{\boldsymbol{q}} S_X \tag{36}$$

The drift entropy flow density

$$\boldsymbol{J}_{st} = \dot{\boldsymbol{X}} S_X \tag{37}$$

The diffusion entropy flow density

$$\boldsymbol{J}_{sd} = -D\nabla_{\boldsymbol{q}} S_X \tag{38}$$

The entropy production density[10-13]

$$\sigma_G = kD\rho(\nabla_{\boldsymbol{q}} \ln \frac{\rho}{\rho_0})^2 = \frac{D}{k\rho}[(\nabla_{\boldsymbol{q}} \ln \rho) S_X - \nabla_{\boldsymbol{q}} S_X]^2 \tag{39}$$

Similarly, we can derive quantum entropy evolution equation in 6N dimensional phase space from the quantum Liouville diffusion equation (6q) and the definition of quantum entropy density operator

$$ih \frac{\partial S_\rho}{\partial t} = [H, S_\rho] + D\nabla_{\boldsymbol{q}}^2 S_\rho + \frac{D}{k\rho}[(\nabla_{\boldsymbol{q}} \ln \rho) S_\rho - \nabla_{\boldsymbol{q}} S_\rho]^2 \tag{35q}$$

where $S_\rho = -k\rho \ln \frac{\rho}{\rho_0}$ or $S_\rho = -k\rho \ln \rho$ is quantum entropy density operator in 6N dimensional phase space.

4.2. 6 and 3 dimensional nonequilibrium entropy evolution equation

Similarly, the nonequilibrium entropy in 6 dimensional phase space can be defined as

$$S_B(t) = -k\int f_1(\boldsymbol{x},t) \ln \frac{f_1(\boldsymbol{x},t)}{f_{10}(\boldsymbol{x})} d\boldsymbol{x} + S_{B0} = \int S_{\mathrm{v}\boldsymbol{p}} d\boldsymbol{x} + S_{B0} \tag{40}$$

where $f_{10}(\boldsymbol{x})$ and $S_{B0}$ are the reduced probability density of the single particle and 6 dimensional entropy for equilibrium states respectively.

The entropy density in 6 dimensional phase space

$$S_{\mathrm{v}\boldsymbol{p}} = -kf_1(\boldsymbol{x},t)\ln\frac{f_1(\boldsymbol{x},t)}{f_{10}(\boldsymbol{x})} \tag{41}$$

or
$$S_B(t) = -k\int f_1(\boldsymbol{x},t)\ln f_1(\boldsymbol{x},t)d\boldsymbol{x} = \int S_{\mathrm{v}\boldsymbol{p}}d\boldsymbol{x} \tag{40a}$$

$$S_{\mathrm{v}\boldsymbol{p}} = -kf_1(\boldsymbol{x},t)\ln f_1(\boldsymbol{x},t) \tag{41a}$$

Differentiating both sides of formula (40) with respect to time $t$ and using the kinetic Eq. (12), we obtain the time rate of change of 6 dimensional nonequilibrium entropy :

$$\frac{\partial S_B}{\partial t} = \int\frac{\partial S_{\mathrm{v}p}}{\partial t}d\boldsymbol{x} = -k\int\left\{[\frac{\partial f_1}{\partial t}(\ln\frac{f_1}{f_{10}}+1)] - \frac{f_1}{f_{10}}\frac{\partial f_{10}}{\partial t}\right\}d\boldsymbol{x}$$

$$= -\int[\nabla_{\boldsymbol{q}_1}\cdot(\boldsymbol{J}_{st}+\boldsymbol{J}_{sd}+\boldsymbol{J}_{\mathrm{v}\boldsymbol{p}}) - \sigma_B]d\boldsymbol{x} \tag{42}$$

and the balance equation of the entropy density in 6 dimensional phase space.

$$\frac{\partial S_{\mathrm{v}\boldsymbol{p}}}{\partial t} == -\nabla_{\boldsymbol{q}_1}\cdot(\boldsymbol{J}_{st}+\boldsymbol{J}_{sd}+\boldsymbol{J}_{\mathrm{v}\boldsymbol{p}}) + \sigma_B \tag{43}$$

Hence we find out the evolution equation of nonequilibrium entropy density in 6 dimensional phase space[12-16]

$$\frac{\partial S_{\mathrm{v}\boldsymbol{p}}}{\partial t} == -\nabla_{\boldsymbol{q}_1}\cdot\left(\boldsymbol{v}S_{\mathrm{v}\boldsymbol{p}} + \boldsymbol{J}_{\mathrm{v}\boldsymbol{p}}\right) - \nabla_{\boldsymbol{p}_1}\cdot\left(\boldsymbol{F}_1 S_{\mathrm{v}\boldsymbol{p}}\right) + D\nabla_{\boldsymbol{q}_1}{}^2 S_{\mathrm{v}\boldsymbol{p}}$$

$$+\frac{D}{kf_1}[(\nabla_{\boldsymbol{q}_1}\ln f_1)S_{\mathrm{v}\boldsymbol{p}} - \nabla_{\boldsymbol{q}_1}S_{\mathrm{v}\boldsymbol{p}}]^2 \tag{44}$$

where $\boldsymbol{v} = \boldsymbol{p}/m$ is the particle velocity, $q_1$ in $\nabla_{\boldsymbol{q}_1}$ represents 3 dimensional coordinate vector $q$ of a particle, i.e. here $\nabla_{\boldsymbol{q}_1} = \nabla_{\boldsymbol{q}}$.

The drift entropy flow density

$$\boldsymbol{J}_{st} = \boldsymbol{v}S_{\mathrm{v}\boldsymbol{p}} \tag{45}$$

The diffusion entropy flow density

$$\boldsymbol{J}_{sd} = -D\nabla_{\boldsymbol{q}_1}S_{\mathrm{v}\boldsymbol{p}} \tag{46}$$

The entropy flow density $\boldsymbol{J}_{\mathrm{v}\boldsymbol{p}}$ in 6 dimensional phase space contributed by the two-particle interaction potential satisfies

$$-\nabla_{\boldsymbol{q}_1}\cdot\boldsymbol{J}_{\mathrm{v}\boldsymbol{p}} = Nk\int(\nabla_{\boldsymbol{q}}\phi)\cdot\left\{\frac{f_1(\boldsymbol{x},t)}{f_{10}(\boldsymbol{x})}\nabla_{\boldsymbol{p}}f_{20}(\boldsymbol{x},\boldsymbol{x}_1) - \nabla_{\boldsymbol{p}}f_2(\boldsymbol{x},\boldsymbol{x}_1,t)\left[1+\ln\frac{f_1(\boldsymbol{x},t)}{f_{10}(\boldsymbol{x})}\right]\right\}d\boldsymbol{x}_1 \tag{47}$$

The entropy production density[12,13]

$$\sigma_B = kDf_1(\nabla_{\boldsymbol{q}_1}\ln\frac{f_1}{f_{10}})^2 = \frac{D}{kf_1}[(\nabla_{\boldsymbol{q}_1}\ln f_1)S_{\mathrm{v}p} - \nabla_{\boldsymbol{q}_1}S_{\mathrm{v}\boldsymbol{p}}]^2 \tag{48}$$

Integrating both side of equation (44) over 3 dimensional momentum space, we obtain the evolution equation of nonequilibrium entropy density in 3 dimensional

space[12-16]

$$\frac{\partial S_V}{\partial t} = -\nabla_{\boldsymbol{q}_1} \cdot (\boldsymbol{C} S_V + \boldsymbol{J}_V) + D\nabla_{\boldsymbol{q}_1}{}^2 S_V$$
$$+\frac{D}{k}\int \frac{1}{f_1}[(\nabla_{\boldsymbol{q}_1} \ln f_1) S_{\mathrm{v}\boldsymbol{p}} - \nabla_{\boldsymbol{q}_1} S_{\mathrm{v}\boldsymbol{p}}]^2 d\boldsymbol{p} \tag{49}$$

where $S_V = \int S_{\mathrm{v}\boldsymbol{p}} d\boldsymbol{p}$ is entropy density in 3 dimensional space . $\boldsymbol{J}_V = \int \boldsymbol{J}_{\mathrm{v}\boldsymbol{p}} d\boldsymbol{p}$ is the entropy flow density in 3 dimensional space contributed by the two-particle interaction potential , $\boldsymbol{C}$ is mean drift velocity of entropy flow.

It should be pointed out here that $\nabla_{\boldsymbol{q}}$ in Eq.(35) is different from $\nabla_{\boldsymbol{q}_1}$ in Eqs.(49),(44). $\boldsymbol{q} = (\boldsymbol{q}_1, \boldsymbol{q}_2, \cdots \boldsymbol{q}_N)$ are the set vectors , but $\boldsymbol{q}_1$ is only one vector $\boldsymbol{q}$ .

Eqs(49)(44) and (35) are three nonequilibrium entropy evolution equations in 3 ,6 and 6*N*-dimensional phase space, which are found out firstly by the author. Their forms are the same. They show that the time rate of change of nonequilibrium entropy density (the term on the left-hand side) is caused together by its drift (the first term on the right-hand side), diffusion (the second term on the right-hand side) and product (the third term on the right-hand side). This means that entropy being an important extensive physical quantity, its density distribution in nonequilibrium statistical thermodynamic systems is always nonuniform, nonequilibrium and changes with time and space. Like mass, momentum and internal energy, the entropy density has both drift and typical diffusion. That is to say, entropy will diffuse from high density region to low density region in nonequilibrium systems. Because entropy represents disorder in the system, nonequilibrium entropy evolution equations of (49)、(44) and (35) represent that local disorder is constantly being produced、drifted and diffused in nonequilibrium systems. Entropy production (or entropy increase)、entropy diffusion、mass diffusion、viscous flow and heat conductivity, all of them are the concrete manifestation of the time's direction or irreversibility of macroscopic processes. Their common microscopic origin is the stochastic diffusion motion of the particles.

Nonequilibrium entropy evolution equations (49)(44)(35) describe the evolution law of nonequilibrium entropy and the evolution process of nonequilibrium systems. They play a central role in nonequilibrium entropy theory. Their importance shall be seen in the following sections 5, 6, 7 and 8, here we give beforehand a brief outline. Entropy production, the third term on the right-hand side, it give a concise statistical formula for the law of entropy increase, i.e. formula (55) (58). Entropy diffusion, the second term on the right-hand side, its presence make us to think that the processes for approach to equilibrium is caused and accomplished by entropy diffusing from its high density region to low density region and finally the distribution of entropy density in all the system reaches to uniformity and the total entropy of the system approaches to maximum. Entropy change contributed by internal two-body interaction potential, the fourth term on the right-hand side of equation (44a), it reveals that internal interaction in a nonequilibrium system can result in entropy decrease or another entropy increase, and their commom expression is formula (96). Obviously, every term among these three terms has itself important physical meaning. As to entropy drift, i.e. entropy flow of open system, the first term on the right-hand side, it is well known in existing theory.

In principle we can solve out the time-space distribution of entropy density from the nonequilibrium entropy evolution equation. But since Eqs.(49)、(44) and (35) are complex nonlinear unclosed partial differential equations, it is difficult to get an rigorous solution.

**5. Formula for entropy production rate**

As is well known, the law of entropy increase, that is the second law of thermodynamics described by entropy, is a fundamental law in nature. It plays an important role not only in physics but also in cosmology、chemistry and biology et el. Although people have done research on it for more than one hundred years since its construction, up to now we do not yet well know the law of entropy increase. What is the microscopic physical basis of entropy production? Which physical parameters and how it changes with? Can it be described by a quantitative concise statistical formula as same as the Boltzmann's entropy formula $S = k\ln W$? This is all along a central problem to be solved in nonequilibrium statistical physics. Recently , the author[15,16] have derived a statistical formula for entropy production rate、namely the law of entropy increase in 6N and 6 dimensional phase space. As its application, we use this formula in calculating and discussing some actual physical topics in the nonequilibrium and stationary states.

According to expressions (39)(48), the entropy production rate in 6N and 6 dimensional phase space are

$$P_G = \frac{d_i S_G}{dt} = \int \sigma_G d\Gamma = kD\int \rho(\nabla_{\boldsymbol{q}} \ln\frac{\rho}{\rho_0})^2 d\Gamma \qquad (50)$$

$$P_B = \frac{d_i S_B}{dt} = \int \sigma_B d\boldsymbol{x}_1 = kD\int f(\nabla_{\boldsymbol{q}_1} \ln\frac{f}{f_0})^2 d\boldsymbol{x}_1 \qquad (51)$$

Expressions (50) (51) in fact are the entropy production rate terms on the right-hand side in nonequilibrium entropy evolution equation (35) (44).Note, $f = f_1$ in expression (51).

It should be pointed out here again that the lower indexes to the operator $\nabla_{\boldsymbol{q}}$ in expressions (50) (51) and the operator $\nabla_{\boldsymbol{q}}^2$ in diffusion term of eqs(35)(44) are coordinate $\boldsymbol{q}$, but not momentum $\boldsymbol{p}$ or other physical parameter. It correctly shows that the entropy production、entropy diffusion and mass diffusion et el all occur in coordinate space, but not momentum space or other physical parameter space.

Now let us find out the concise statistical formula for the entropy production rate, i.e. the law of entropy increase from expressions (50) and (51).

We begin at first from formula (50). Being similar to the definition of strain or elongation percentage[29] $\varepsilon = \ln(l/l_0)$ of solid materials under the action of complex stress, we can define a new physical parameter of nonequilibrium system, that is the departure percentage from equilibrium of the ensemble probability density of nonequilibrium system in $6N$ dimensional phase space as

$$\theta = \ln\frac{\rho}{\rho_0} \approx \frac{\Delta\rho}{\rho_0} \qquad (52)$$

Using the number density of micro-states $\omega_0$ and $\omega$ for equilibrium and nonequilibrium states, which satisfy the relations $\omega_0 = W_0\rho_0$ and $\omega = W\rho$ respectively, then (52) changes to

$$\theta = \ln\frac{\omega}{\omega_0} - \ln\frac{W}{W_0} \simeq \frac{\Delta\omega}{\omega_0} - \frac{\Delta W}{W_0} \tag{53}$$

It also can be defined as the departure percentage from equilibrium of the difference between the number density and the total number of micro-states of nonequilibrium system in 6N dimensional phase space. Because the number of micro-states represents the degree of disorder, $\theta$ also can be understood as the departure percentage from equilibrium of the degree of disorder density of nonequilibrium system. For brevity, we call $\theta$ as the departure percentage from equilibrium in the following. Please notice here, the last approximate equalities in (52) and (53) are valid only when $\frac{\Delta\rho}{\rho_0} << 1$ and $\frac{\Delta\omega}{\omega_0} << 1$.

The space gradient of the departure percentage from equilibrium is

$$\nabla_q\theta = \nabla_q \ln\frac{\rho}{\rho_0} = \nabla_q \ln\frac{\omega}{\omega_0} \tag{54}$$

Substituting (54) into (50), then the entropy production rate in 6N dimensional phase space changes to[15,16]

$$P_G = kD\int \rho(\nabla_q \ln\frac{\omega}{\omega_0})^2 d\Gamma = kD\int \rho(\nabla_q\theta)^2 d\Gamma = kD\overline{(\nabla_q\theta)^2}$$

That is $\qquad P_G = kD\overline{(\nabla_q\theta)^2} \geq 0 \qquad (55)$

Where $\overline{(\nabla_q\theta)^2} = \int \rho(\nabla_q\theta)^2 d\Gamma$ is the average value of the square of the space gradient of the departure percentage from equilibrium.

Corresponding to formula (55), the formula for quantum entropy production rate can be obtained from equation (35q)

$$P_{qG} = ih\frac{\partial_i S}{\partial t} = kDTr[\rho(\nabla_q\theta_\rho)^2] \geq 0 \tag{55q}$$

Similarly, we can define the departure percentage from equilibrium of nonequilibrium system in 6 dimensional phase space as

$$\theta_b = \ln\frac{f}{f_0} = \ln\frac{\omega_b}{\omega_{b0}} - \ln\frac{W_b}{W_{b0}} \tag{56}$$

Where $\omega_{b0}$ and $\omega_b$ are the number density of micro-states for equilibrium and nonequilibrium states in 6 dimensional phase space respectively. The space gradient of the departure percentage from equilibrium is

$$\nabla_{q_1}\theta_b = \nabla_{q_1} \ln\frac{f}{f_0} = \nabla_{q_1} \ln\frac{\omega_b}{\omega_{b0}} \tag{57}$$

Substituting (57) into (51), then the entropy production rate in 6 dimensional phase space

changes to[15,16]

$$P_B = kD\int f(\nabla_{q_1} \ln\frac{\omega_b}{\omega_{b0}})^2 d\boldsymbol{x}_1 = kD\int f(\nabla_{q_1}\theta_b)^2 d\boldsymbol{x}_1 = kD\overline{(\nabla_{q_1}\theta_b)^2}$$

That is $$P_{\boldsymbol{B}} = kD\overline{(\nabla_{q_1}\theta_b)^2} \geq 0 \qquad (58)$$

Where $\overline{(\nabla_{q_1}\theta_b)^2} = \int f(\nabla_{q_1}\theta_b)^2 d\boldsymbol{x}_1$ is the average value of the square of the space gradient of the percentage departure from equilibrium.

When the system is in statistically independent state, that is $\rho(\boldsymbol{X},t) = f(\boldsymbol{x}_1,t)f(\boldsymbol{x}_2,t)\cdots f(\boldsymbol{x}_N,t)$**,** then

$$P_G = NkD\overline{(\nabla_{q_1}\theta_b)^2} = NP_B \geq 0 \qquad (59)$$

Formulas (55) and (58) are the concise statistical formulas for entropy production rate in 6N and 6 dimensional phase space, which we derive from nonequilibrium entropy evolution equations (35)(44). This is also the quantitative concise statistical formula for the law of entropy increase. It shows that the entropy production rate *P* equals to the product of diffusion coefficient *D*、the average value of the square of space gradient of the departure percentage from equilibrium $\overline{(\nabla_{\boldsymbol{q}}\theta)^2}$ and Boltzmann constant *k*. Obviously, for nonequilibrium ($\theta \neq 0$)、spatially inhomogeneous ($\nabla_{\boldsymbol{q}}\theta \neq 0$) system with stochastic diffusion (*D*≠0), the entropy always increase ($P_G > 0$). Conversely, for equilibrium system ($\theta = 0$) or that nonequilibrium but spatially homogeneous ($\nabla_{\boldsymbol{q}}\theta = 0$)system or that system only with deterministic but no stochastic motion (*D*=0), there all are no entropy production ($P_G = 0$). It should be pointed out here that the entropy in statistical thermodynamic system is possible to produce only when the motion of its internal particles has the stochastic diffusion form; the entropy does not increase with time for that system whose internal particles move only with deterministic but not stochastic form. In other words, the stochastic diffusion motion of particles is the microscopic origin of entropy production and manifests the dissipative character of entropy production. It is also easy to see that the space gradient of the departure percentage from equilibrium of the number density of micro-states、namely the spatially inhomogeneous departures from equilibrium of the system, which is important more than the departure percentage from equilibrium for determining entropy production, is a microscopic basis of entropy production. Combining these two we may have the idea that the spatially stochastic and inhomogeneous departure from equilibrium of the number density of micro-states is the microscopic physical basis of the macroscopic entropy production in nonequilibrium system.

Formulas (55) and (58) clearly tell us that the entropy production rate of a nonequilibrium physical system is only determined by two physical parameters: diffusion coefficient *D* and the departure percentage from equilibrium $\theta$ (not including the known constant *k*). It can be used in calculating the entropy production rate of the actual physical topics when *D* and $\theta$ are known. The diffusion coefficient *D* can be calculated and measured from experiment. The departure percentage from equilibrium $\theta$, a new defined physical parameter, its physical meaning is clear, whose introduction not only makes the formula for the entropy production rate simple and clear, but also makes it can play the role of an physical parameter to describe quantitatively how far a nonequilibrium system is from equilibrium as if the strain describes the deformation of the solids. If we compare the Boltzmann's entropy formula with the formula of entropy production rate (55) (58), it is easy to see, the former shows that the macroscopic entropy of a physical system is determined in terms of the total number of micro-states, and the latter demonstrates that the macroscopic entropy production of a nonequilibrium physical system is caused by spatially stochastic and inhomogeneous departure from equilibrium of the number density of micro-states. Both the concise statistical formulas for macroscopic entropy are related to the number of micro-states and provide a bridge between the macroscope and microscope.

It can be seen from the formula (55)(58)for the law of entropy increase that the arrow of time for an isolated system always points to the direction of entropy increase , and the arrow rate is determined by the entropy increase rate.

By the way, from formula (50) we can also obtain the theorem of minimum entropy production[12,13].

### 5.1 Entropy production in nonequilibrium state

Now let us use the statistical formula (58)in calculating and discussing the following three nonequilibrium physical topics. For the convenience of calculation we only discuss one dimensional problems. If neglecting the interaction between two particles, the term including the probability density of the two particles vanishes, then the kinetic equation(12) for the probability density of the single particle changes into a standard Fokker-Planck equation. In this and next sections our discussion is based on this equation.

#### 5.1.1 Free expansion of gas

The entropy increase of this topic is a classical example in equilibrium statistical thermodynamics. However, equilibrium statistical thermodynamics only gives the entropy production for final equilibrium state, but cannot for nonequilibrium state.

According to the idea of nonequilibrium statistical physics, the free expansion process of gas may be regarded as a self-diffusion process of gas. For the sake of calculating convenience, assume that the container is a cylinder, whose cross section is unit and whose length is as long as concentration changes of gas near boundaries are negligible. Thus the diffusion of gas in the cylinder can be reduced to one dimensional problem [30]. The gas is initially confined to the left half ($q$<0) of cylinder by a partition, the right half ($q$>0) is empty. After the removal of the partition the gas will freely diffuse and expand into the right half of cylinder. Now let $t$ denote the time that the gas diffuses, $q$ denote the position in cylinder, $C (q, t)$ denote the concentration of gas at time t and position $q$, then from solving one dimensional diffusion equation we obtain[30].

$$C(q,t) = \frac{C_0}{2}\left[1 - erf(q/2\sqrt{Dt})\right] \tag{60}$$

Obviously, formula (60) satisfies the initial condition

$$C = C_0 \text{ for } q < 0, \quad \text{and } C = 0 \text{ for } q > 0 \quad \text{at } t = 0$$

When $t\to\infty$, $C = \dfrac{C_0}{2}$ for any $q$ everywhere, the gas approaches to final equilibrium state.

where $C_0$ is initial concentration of gas, erf ( $q/2\sqrt{Dt}$ ) is Gauss error function.

From (60) we obtain the probability to find the gas particle at time $t$ and position $q$

$$w(q,t)=\frac{1}{2}[1-erf(q/2\sqrt{Dt})]$$

The probability density is

$$f(q,t)=\frac{dw}{dq}=\frac{1}{2\sqrt{\pi Dt}}\exp(-\frac{q^2}{4Dt}) \qquad (61)$$

Obviously, $f(q,t)$ satisfies the normalization condition $\int_{-\infty}^{\infty} f(q,t)\,dq = 1$ .

The equilibrium probability density is

$$f_0(q)=cons\tan t \qquad (62)$$

Substituting (61) (62) into (56) (58), we obtain the expressions of $\theta_b$ and the entropy production rate of gas diffusion and expansion at time $t$

$$P_B = kD\int_{-\infty}^{\infty} f\left(\frac{\partial \theta_b}{\partial q}\right)^2 dq = kD\int_{-\infty}^{\infty} f(q,t)\left[\frac{\partial}{\partial q}\ln\frac{f(q,t)}{f_0(q)}\right]^2 dq = \frac{k}{t} \qquad (63)$$

The entropy production at time $t$ is

$$\triangle_i S = \int_0^t P_B\,dt = k\int_0^t \frac{dt}{t} \qquad (64)$$

We see here immediately that both $P_B$ and $\Delta_i S$ become infinite at $t=0$ from (63)(64), this is due to (60) that the concentration is infinitive at $t=0$ and $q=0$. In fact, for $t=0$, the system does not change, hence the entropy production $\Delta_i S=0$ and the entropy production rate $P_B$ should be finite. In order to satisfy this initial condition, (63)(64) must be modified into

$$P_B=\frac{k}{t_0+t} \qquad (65)$$

$$\Delta_i S = k\ln\frac{t_0+t}{t_0} \qquad (66)$$

The constant $t_0$ can be thought as a time that the gas has diffused and expanded before $t=0$ inside the left half of cylinder. From equilibrium statistical thermodynamics [4] we know that for $t=0$ the entropy of gas inside the volume $V_1$ of the left half of cylinder is $S_1 \sim k\ln V_1 = k\ln \gamma t_0$, here $\gamma$ is another constant, its physical meaning can be thought as a average volume that the gas diffuse and expands inside cylinder for unit time, hence $t_0=V_1/\gamma$. Substituting this into (65)(66), then the expressions of the entropy production rate、its time rate and entropy production of a single gas particle that the gas diffuses and expands for time $t$ can be written

$$P_B=\frac{\gamma k}{V_1+\gamma t}\geq 0 \qquad (67)$$

$$\frac{\partial P_B}{\partial t} = -\frac{\gamma^2 k}{(V_1 + \gamma t)^2} \le 0 \tag{68}$$

$$\Delta_i S = k \ln \frac{V_1 + \gamma t}{V_1} \ge 0 \tag{69}$$

If $t_f$ is the time that the system approaches to final equilibrium state, than the entropy production rate、its time rate and entropy production of N gas particle at this time are

$$P_B{}^N = \frac{N\gamma k}{V_1 + \gamma t_f} \approx 0 \tag{70}$$

$$\frac{\partial P_B}{\partial t} = -\frac{N\gamma^2 k}{(V_1 + \gamma t_f)^2} \approx 0 \tag{71}$$

$$\Delta_i S^N = Nk \ln \frac{V_1 + \gamma t_f}{V_1} = Nk \ln \frac{V_2}{V_1} = Nk \ln 2 \tag{72}$$

Where $V_2 = V_1 + \gamma t_f = 2V_1$ is the total volume of the left and right two halves of cylinder. The last equalities in (70)(71) are valid owing to that $t_f$ may be regarded very large. The formula (72) just is the well known result of equilibrium statistical thermodynamics [4]. However, the formula (67)(68) (69) changing with time for the entropy production rate、its time rate and entropy production of nonequilibrium state cannot be given by equilibrium statistical thermodynamics. Equations (68)(71) is the expression of the theorem of minimum entropy production for this topic.

5.1.2 Brownian motion

Brownian motion is a typical topic in nonequilibrium statistical physics. If $f(q,t)$ denotes the probability density to find one dimensional Brownian particle in position between $q$ and $q+dq$ at time $t$, then from solving one dimensional Fokker-Planck equation [31] we have

$$f(q,t) = [\pi a(t)]^{-1/2} \exp\{-[q - b(t)]^2 / a(t)\} \tag{73}$$

where $a(t) = a_m(1 - e^{-2\beta t}) + a_0 e^{-2\beta t}$

$$a_m = \frac{2D}{\beta} \qquad b(t) = b_0 e^{-\beta t}$$

$\beta$ is the friction constant, $a_0$ and $b_0$ are two initial constant.

The equilibrium probability density is

$$f_0(q) = (\pi a_m)^{-1/2} \exp(-\frac{q^2}{a_m}) \tag{74}$$

Obviously, $f(q,t)$ satisfies the normalization condition $\int_{-\infty}^{\infty} f(q,t)dq = 1$ .

Substituting (73)(74) into (56)(58), we obtain the expressions of $\theta_b$ and the entropy production rate of the system with Brownian motion at time $t$

$$P_B = kD \int_{-\infty}^{\infty} f(\frac{\partial \theta_b}{\partial q})^2 dq$$

$$=k\beta\left[\frac{2b^2(t)}{a_m}+\frac{a(t)}{a_m}+\frac{a_m}{a(t)}-2\right]\geq 0 \tag{75}$$

The time rate of entropy production rate at time $t$

$$\frac{\partial P_B}{\partial t}=-2k\beta^2\left[\frac{2b^2(t)}{a_m}+\frac{a(t)}{a_m}+\frac{a^2{}_m}{a^2(t)}-\frac{a_m}{a(t)}-1\right]\leq 0 \tag{76}$$

where

$$\frac{a(t)}{a_m}+\frac{a^2{}_m}{a^2(t)}-\frac{a_m}{a(t)}-1=\left[\sqrt{\frac{a(t)}{a_m}}-\sqrt{\frac{a_m}{a(t)}}\right]^2+\left[\frac{a_m}{a(t)}-1\right]^2$$

The entropy production at time $t$ is

$$\Delta_i S=\int_0^t P_B\,dt=\frac{kb_0^{\ 2}}{a_m}(1-e^{-2\beta t})+\frac{k}{2}(e^{-2\beta t}-1)+\frac{ka_0}{2a_m}(1-e^{-2\beta t})$$
$$+\frac{k}{2}\ln\frac{a_m+(a_0-a_m)e^{-2\beta t}}{a_0}\geq 0 \tag{77}$$

From (75)(76)(77)the entropy production rate、its time rate and entropy production at initial state $t=0$ are

$$\begin{cases} P_B=k\beta\left(\dfrac{2b_0^{\ 2}}{a_m}+\dfrac{a_0}{a_m}+\dfrac{a_m}{a_0}-2\right)>0 \\ \dfrac{\partial P_B}{\partial t}=-2k\beta^2\left[\dfrac{2b_0}{a_m}+\dfrac{a_0}{a_m}+\dfrac{a^2{}_m}{a^2{}_0}-\dfrac{a_m}{a_0}-1\right]<0 \\ \Delta_i S=0 \end{cases} \tag{78}$$

The entropy production rate、its time rate and entropy production at final equilibrium $t=t_f\geq\beta^{-1}$ state are

$$\begin{cases} P_B=0 \\ \dfrac{\partial P_B}{\partial t}=0 \\ \Delta_i S=\dfrac{kb_0^{\ 2}}{a_m}+\dfrac{k}{2}\left(\dfrac{a_0}{a_m}-\ln\dfrac{ea_0}{a_m}\right)>0 \end{cases} \tag{79}$$

It can be seen from (75)—(79) that the entropy production rate $P_B>0$、its time rate $\frac{\partial P_B}{\partial t}<0$ and entropy production $\Delta_i S=0$ at initial state, $P_B=0$、$\frac{\partial P_B}{\partial t}=0$ and $\Delta_i S>0$ at final state, and $P_B>0$、$\frac{\partial P_B}{\partial t}<0$ and $\Delta_i S>0$ for any other time ($0<t<\infty$). Equation (76) is the expression of the theorem of minimum entropy production for Brownian motion. These results are reasonable in physics.

We can see in above topics that reducing (50) (51) to (55) (58) does not make the calculation more simple, it seems as if that this reduction has no any advantage. However, from the qualitative discussion of the following topic, the introduction of $\theta$ not only makes the formula simple and clear, but also directly leads to a better understanding of physical problem.

5.1.3 Deformation and fracture of solids

Solids undergo elastic and plastic deformation under the action of stress. The former is reversible, the latter is irreversible. Up to now the entropy change in the latter process is not clear. Here we are also difficult to give a quantitative result of entropy production as above, but we want to know from (55) (58) what new qualitative physical inference can be drawn and confirmed.

According to (55)(58), the space gradient of the departure percentage from equilibrium of the number density of micro-states $\nabla_q \theta$, that is in essence the spatially inhomogeneous departure from equilibrium of the micro-structure, is a necessary premise for the entropy production in nonequilibrium irreversible processes. From this it can be drawn a new physical inference: the change of micro-structure within system during an irreversible processes is inhomogeneous. We use this inference to judge the elastic deformation: because this process is reversible and deformed homogeneously, there should be no entropy production. Experiments really prove that the entropy does not change for purely shearing elastic deformation of solids. We use this inference to judge plastic deformation: because this process is irreversible and the entropy increases. The change of the micro-structure of crystal should be inhomogeneous. Experiments surely confirms that for even very pure plastically deformed monocrystal, the slip lines on its surface always concentrate into slip bands[32], and the distribution in space of its internal glide dislocations is always inhomogeneous[32]. The experimental phenomenon of this slip inhomogeneity was very difficult to understand for a long time, but now it becomes to be clear at a glance in terms of the viewpoint of the new inference drawn from the formula (55)(58) for the entropy production.

In fact, not only plastic deformation, but also the fracture processes, as a irreversible processes occurring in same time or a little later, the change of the micro-structure of solids is also inhomogeneous. Its typical manifestation is that the distribution of microcrack nucleation, growth and propagation, which directly leads to fracture, is always inhomogeneous[29,32] in space. This inhomogeneity of the microcrack evolution also confirms the new inference drawn from formula (55)(58) .

Thus it can be seen, for all those more complex evolutional system, if you ask why the change of their microstructure is inhomogeneous, we can also give a qualitative unified interpretation from this new inference.

5.2 Entropy production in stationary state

The common character for stationary and equilibrium states of macroscopic system is that their macro-states does not change with time. The difference between them is that a macroscopic current exists in the stationary state, but does not in equilibrium state. If we use entropy language, their characters can be described as: The entropy production and entropy flow are present in the stationary state, but equal to zero in the equilibrium state. The total entropy of both kinds of the systems does not change with time.

5.2.1 The formula in stationary state

Now let us derive a special statistical formula for entropy production rate in the stationary state from one dimensional Fokker-Planck equation. According their definition, we have $\frac{\partial f(q)}{\partial t}=-\frac{\partial J}{\partial q}=0$ in both stationary and equilibrium states, and the probability current [33]

$$J=K(q)f(q)-D\frac{\partial f(q)}{\partial q} \tag{80}$$

is constant. For equilibrium state, $J=0$, the probability density of the system is

$$f_0(q)=n_0\exp\left[\frac{1}{D}\int_0^q K(q)dq\right]=n_0\exp[-\phi(q)] \tag{81}$$

Where $\phi(q)=-\frac{1}{D}\int_0^q K(q')dq'$

For stationary state, $J\neq 0$, the probability density of the system is

$$\begin{aligned} f_{st}(q)&=n\exp[-\phi(q)]-\frac{J}{D}\exp[-\phi(q)]\int_0^q\exp[\phi(q')]dq' \\ &=n(q)\exp[-\phi(q)] \end{aligned} \tag{82}$$

where $n_0$ and $n$ are the normalization constants. When $J=0$, the system return to equilibrium state (81) from stationary state (82).

Substituting (81) (82) into (56) (58), we obtain the expressions of $\theta_b$ and the entropy production rate in the stationary state

$$\begin{aligned} P_B&=kD\int_0^L f_{st}(q)\left(\frac{\partial\theta_b}{\partial q}\right)^2dq=kD\int_0^L f_{st}(q)\left[\frac{\partial}{\partial q}\ln\frac{f_{st}(q)}{f_0(q)}\right]^2dq \\ &=\frac{kJ^2}{D}\int_0^L\frac{dq}{f_{st}(q)}=kJ\left[-\phi(L)+\ln\frac{f_{st}(0)}{f_{st}(L)}\right] \end{aligned}$$

That is
$$P_B=kJ\left[-\phi(L)+\ln\frac{f_{st}(0)}{f_{st}(L)}\right]\geq 0 \tag{83}$$

In obtaining the last equality in (83) we have used the following integral result

$$\begin{aligned} \int_o^L\frac{dq}{f_{st}(q)}&=\int_o^L\frac{\exp[\phi(q)]dq}{n-\frac{J}{D}\int_o^q\exp[\phi(q')]dq'} \\ &=-\frac{D}{J}\left[\phi(L)+\ln\frac{n\exp[-\phi(L)]-\frac{J}{D}\exp[-\phi(L)]\int_o^L\exp[\phi(q)]dq}{n}\right] \end{aligned}$$

Where $L$ is the spatial length $[O, L]$, in which the $J$ flows. $f_{st}(L)$ and $f_{st}(O)$ are the probability density at boundaries $q=L$ and $q=O$.

Formula (83) is just the special statistical formula for entropy production rate in one dimensional stationary state. It shows that the entropy production rate in the stationary state is proportional to its probability current. When $J=0$, then $P_B=0$. This means that the entropy production rate is present only in that stationary state with nonzero probability current. All those stationary states with zero probability current in fact are equilibrium states, their entropy production rate only can be zero. It should be also pointed out here although formula (83) is one dimensional, it is easy to generalize it to two and three dimensional stationary states. How to calculate the actual expressions of $J$、$f_{st}$ and $P_B$? Please see the following two topics with and without macroscopic external force.

5.2.2 Directed atomic diffusion

The directed diffusion of atom under the action of constant external force is a typical stationary state system. When there is no external force, the diffusion process of atoms obeys the diffusion equation, in which no directed motion is present. While the atomic system is subjected to a constant external force $F$, the directed diffusion is produced. And the average directed diffusion velocity of the atoms is [34]

$$K = V = \frac{DF}{kT} \tag{84}$$

Thus the transport process of the atomic system can be described by the Fokker-planck equation[34]. Substituting (84) into (82) and using the normalization condition $\int_0^L f_{st}(q)dq = 1$, we obtain the probability density

$$f_{st}(q) = (n - \frac{J}{V})\exp(\frac{\text{qV}}{D}) + \frac{J}{V} \tag{85}$$

the normalization constant

$$n = \left(\frac{V - JL}{D}\right)\left[\exp\left(\frac{VL}{D}\right) - 1\right]^{-1} + \frac{J}{V} \tag{86}$$

The Probability current of the system in the stationary state

$$J = \bar{n}V \tag{87}$$

Where $\bar{n}$ is the average particle number per unit length, it can be defined as

$$\bar{n} = \int_O^L n(q) f_{st}(q)dq \tag{88}$$

Substituting (82) (85) into (88) and jointly solving with (86), we obtain the expression of probability current

$$J \approx \frac{V}{aL} = \frac{DF}{aLkT} \tag{89}$$

Substituting (89) into (85) (86), then substituting them into (83) and using (84), we obtain the entropy production rate for the directed diffusion of atomic system in the stationary state

$$P_B = kJ\left[\frac{FL}{kT} - \ln\frac{f_{st}(L)}{f_{st}(O)}\right] > 0 \tag{90}$$

where

$$\ln\frac{f_{st}(L)}{f_{st}(O)} = \ln\left\{\frac{\frac{\gamma F}{kT}\exp\left(\frac{FL}{kT}\right) + \frac{1}{L}\left[\exp\left(\frac{FL}{kT}\right) - 1\right]}{\frac{\gamma F}{kT} + \frac{1}{L}\left[\exp\left(\frac{FL}{kT}\right) - 1\right]}\right\}$$

The above two constants are: $a$=1.61; $\gamma$=0.38. It can be seen from (89)(90) that the entropy production rate is determined by the external force and the diffusion coefficient. When the external force *F=0*, or the diffusion coefficient *D=0,* then the probability current $J = 0$, so that the entropy production rate $P_B = 0$.

5.2.3 Molecular Motor

Molecular motor plays important role in life process. It can convert chemical energy into mechanical energy effectively and directly. Comparing with the above directed diffusion of atom caused by external force, the directed motion of molecular motor performs in the absence of any macroscopic external force. What is the mechanism of this directed motion? It has stimulated interest of many biologists and physicists, and many different models were proposed. Here we use a periodically rocked model[35,36] of Brownian motor to calculate the entropy production rate of the molecular motor. In this model, molecular motor are regarded as a Brownian particle, which are driven by three forces. That is: the spatial unsymmetrical periodical potential、the temporal periodical force and a Gaussian white thermal noise. The two formers originate from the motor system, the third come from environment. If $q$ denotes the state of molecular motor, then we have

$$\dot{q} = -\frac{\partial}{\partial q}\left[U(q) - qV(t)\right] + \eta(t) = K(q) + \eta(t) \tag{91}$$

We take the spatial unsymmetrical periodical potential $U(q) = -\frac{1}{2\pi}\left[\sin(2\pi q) + \frac{1}{4}\sin(4\pi q)\right]$, temporal periodical force $V(t) = A\sin(wt)$, and $\eta(t)$ is a Gaussian white noise. As we hope that the spatial average of $\frac{\partial}{\partial q}U(q)$ as well as the time average of $V(t)$ are zero. Because $V(t)$ is time-dependent, the Fokker-Planck equation being equivalent to the Langevin equation (91) does not exist stationary solution. For this reason we may take $w \ll 1$, so that $V(t)$ changes with $t$ very slowly. Thus the system exist quasi-stationary state, and its probability current can be approximately determined by (80). Substituting $K(q) = \cos(2\pi q) + \frac{1}{2}\cos(4\pi q) + A\sin(wt)$ into (80)(82), using boundary condition $f_{st}(L) = f_{st}(O)$ and normalization condition $\int_O^L f_{st}(q)dq = 1$, then gives the probability current of the system in the quasi-stationary state in the expression[36]

$$J = D\left\{\left[1 - \exp\left(-\frac{LA}{D}\sin wt\right)\right]^{-1}\int_O^L dq\int_O^L dq'\exp\left[\phi(q,t) - \phi(q',t)\right]\right.$$

$$\left. - \int_O^L dq\int_O^q dq'\exp\left[\phi(q't) - \phi(q,t)\right]\right\}^{-1} = DJ_d \tag{92}$$

where $\phi(q,t) = \left[U(q) - qA\sin(wt)\right]/D$, spatial period ***L***=1.

The probability density of quasi-stationary state

$$f_{st}(q) = \frac{J}{D}\exp\left[-\phi(q,t)\right]\left\{\left[1 - \exp\left(-\frac{LA}{D}\sin wt\right)\right]^{-1}\int_O^L \exp\left[\phi(q,t)\right]dq\right.$$

$$-\int_O^q \exp[\phi(q',t)]dq' \Big\} \quad (93)$$

Substituting $V$、$\phi(L)$、$J$ and $f_{st}(L) = f_{st}(O)$ into (83), we obtain the entropy production rate for molecular motor in quasi-stationary state

$$P_B = \frac{kJLA\sin(wt)}{D} = kJ_d LV(t) \quad (94)$$

Expression (94) shows that the entropy production rate $P_B$ is proportional to $J/D=J_d$ and $V(t) = A\sin(wt)$. Because $J_d$ is also changes with $V(t)$ and $D$ complicatedly, $P_B$ is a complicated nonlinear function of $V(t)$ and $D$. The numerical data for the entropy production rate $P_B$ vs driving amplitude $A$ and diffusion coefficient $D$ shows that $P_B$ first increases、then decrease with $A$ after a maximum; and $P_B$ increases with $D$ and tends to saturation. $P_B$ =0 both at $A$=0 and $D$=0.

Because the entropy production rate $P_B > 0$, we can conclude that the efficiency of molecular motor is still smaller than 100% though it is very high. That is, the second law of thermodynamics is also universal for molecular motor.

It should be also pointed out here, although the calculated actual topics in this paper are one dimensional, the statistical formula (58) (55) in fact can be applied to calculate the entropy production rate in 2、3、6 and 6N dimensional space. The only difference is that the calculation is more complex.

5.3 A short conclusion and discussion

Starting from the nonequilibrium entropy evolution equation, first we defined a new physical parameter, the departure percentage from equilibrium, which can quantitatively describes how far a nonequilibrium system is from equilibrium. Then we derived a concise statistical formula for entropy production rate、namely the law of entropy increase in 6N and 6 dimensional phase space. It shows that the spatially stochastic and inhomogeneous departure from equilibrium of the number density of micro-states is the microscopic physical basis of the macroscopic entropy production in nonequilibrium system. From this formula, we derived a formula in stationary state which proves that the entropy production rate in the stationary state is proportional to its probability current. Furthermore, we present some actual expressions of entropy production、their first and second time rate in the nonequilibrium states, and the entropy production rate in the stationary states.

## 6. Entropy change from internal interaction

The law of entropy increase[1-3,28] shows that if an isolated system is not in a statistical equilibrium state, its macroscopic entropy will increase with time, until ultimately the system reaches a complete equilibrium state where the entropy attains its maximum value. According to the inference of this law, the universe is as isolated system, it also ought to degrade into a complete statistical equilibrium state, i.e.the so-called heat death state.However, `the real world is another scene.Everywhere there is order and structure： stars, galaxies, plants and animals etc. They are always incessantly evolving. When the law of entropy increase occupies a dominant position, why can an isolated system create order structure? Does the entropy of an isolated system always only increase and never decrease? Why can the life from nothing to some thing with simple atoms and molecules organize itself into a whole? Why can they resist the law of entropy increase and produce self-organizing structure? Whether or not because they are also governed by a power of some unknown entropy decrease inherent in the system? If yes, what is its dynamical mechanism? And what is its mathematical expression? What is the difference in mathematical form and microscopic physical foundation between the formulas for this entropy decrease and the law of entropy increase?

In the late half of the twentieth century, the publications of the theory of dissipative structures[37], synergetics[31] and the hypercycle[38] marked an important progress of quantitative theories in self-organization. However, these theories, including the formal entropy theory[37] decomposing the entropy change into the sum of the entropy flow and the entropy production, discuss only the problems of open system but not isolated system. From the point of view of exploring that what isolated system can appear the entropy decrease to be a match for the law of entropy increase, they all have no relation.

The main results in this section are as follows.[18]

In order to investigate entropy decrease, we write again nonequilibrium entropy evolution equation (44) in 6 dimensional phase space as follows (now assume $\boldsymbol{F}=0$)

$$\frac{\partial \mathrm{S}_{\mathrm{v}\boldsymbol{p}}}{\partial t}=-\nabla_{\boldsymbol{q}_l}\cdot(\boldsymbol{v}\mathrm{S}_{\mathrm{v}\boldsymbol{p}})+D\nabla^2_{\boldsymbol{q}_l}\mathrm{S}_{\mathrm{v}\boldsymbol{p}}+\frac{D}{kf_1}\Big[(\nabla_{\boldsymbol{q}_l}\ln f_1)\mathrm{S}_{\mathrm{vp}}-\nabla_{\boldsymbol{q}_l}\mathrm{S}_{\mathrm{v}\boldsymbol{p}}\Big]^2+\lambda(\phi) \qquad (44a)$$

The entropy density change rate $\lambda(\phi)=-\nabla_{\boldsymbol{q}_l}\cdot\boldsymbol{J}_{\mathrm{v}\boldsymbol{p}}$ contributed by the two-particle interaction potential originates from the first term on the right-hand side of the BBGKY diffusion equation (12),

Just this $\lambda(\phi)$ results in the entropy decrease. In other words, the entropy decrease probably appears in nonequilibrium system with internal attractive interaction. Before proving this proposition, let us consider the relation between a famous kinetic equation — the Boltzmann equation and the BBGKY diffusion equation (12). In fact the former is a variety of the latter (when there is no diffusion term) and is applied to describe the motion of a dilute gas system with internal short-range repulsive interaction or collision. More concretely saying, that is to change the first term with $f_2(\boldsymbol{x},\boldsymbol{x}_1,t)$, the two particle interaction term, on the right hand side of equation (12) into collision term only with $f_1(\boldsymbol{x},t)$ of dilute gas particle. All other terms do not change. As a result, the entropy of this nonequilibrium gas system increases due to the existence of short-range repulsive interaction or collision. In short, the entropy increase in the system described by the Boltzmann equation originates from short-range repulsive interaction or collision between internal particles. This enlightens us: when the interaction force among internal particles is attractive but not repulsive, the entropy decrease probably appears in nonequilibrium system. In view of that all present closed kinetic equations are changed from the unclosed BBGKY equation (12), and up to now we did not know what closed kinetic equation should be applied to describe a nonequilibrium statistical system with internal attractive interaction, so the kinetic equation to describe galactic dynamics is yet the BBGKY equation(12) [39]. This is the reason why our discussion is based on the BBGKY diffusion equation (12). Thus, our proof reduces to: when the internal two-particle interaction potential $\phi<0$, the time rate of change of entropy $R(t)$ described by the fourth term on the right-hand side of equation(44a) should be negative. That is $R(t)=\int\lambda(\phi)d\boldsymbol{x}<0$. According to expression (47), we obtained its mathematical

expression

$$R(t)=\int\lambda(\phi)d\boldsymbol{x}\backslash=Nk\int(\nabla_q\phi)\cdot\left\{\frac{f_1(\boldsymbol{x},t)}{f_{10}(\boldsymbol{x})}\nabla_p f_{20}(\boldsymbol{x},\boldsymbol{x}_1)-\nabla_p f_2(\boldsymbol{x},\boldsymbol{x}_1,t)\left[\ln\frac{f_1(\boldsymbol{x},t)}{f_{10}(\boldsymbol{x})}\right]\right\}d\boldsymbol{x}d\boldsymbol{x}_1$$

$$=Nk\int f_2(\boldsymbol{x},\boldsymbol{x}_1,t)(\nabla_q\phi)\cdot\left\{\nabla_p\left[\ln\frac{f_1(\boldsymbol{x},t)}{f_{10}(\boldsymbol{x})}\right]-\frac{f_{20}(\boldsymbol{x},\boldsymbol{x}_1)}{f_2(\boldsymbol{x},\boldsymbol{x}_1,t)}\nabla_p\left[\frac{f_1(\boldsymbol{x},t)}{f_{10}(\boldsymbol{x})}\right]\right\}d\boldsymbol{x}d\boldsymbol{x}_1 \quad (95)$$

Now let us simplify expression (95) and investigate its physical meaning. Similar to introducing $\theta_1$ into formula (55)(58), here we also define $\theta_2=\ln(f_2/f_{20})=-\ln(\omega_2/\omega_{20})$ as the departure percentage from equilibrium of nonequilibrium system in 12 dimensional phase space. Substituting $\theta_1$ and $\theta_2$ into last line of expression (95) and through some operation, we have

$$R(t)=Nk\int f_2(\boldsymbol{x},\boldsymbol{x}_1,t)(\nabla_q\phi)\cdot(\nabla_p\theta_1)\left[1-e^{-(\theta_2-\theta_1)}\right]d\boldsymbol{x}d\boldsymbol{x}_1$$

$$=Nk\overline{(\nabla_q\phi)\cdot(\nabla_p\theta_1)\left[1-e^{-(\theta_2-\theta_1)}\right]} \quad (96)$$

This is the statistical expression for the time rate of change of macroscopic entropy caused by the internal two-body interaction in nonequilibrium system. It shows that the macroscopic entropy change rate $R(t)$ of nonequilibrium system equals to the average value of the product of internal interaction force $\nabla_q\phi$, momentum space gradient $\nabla_p\theta_1$ of the departure percentage from equilibrium and the difference of the percentage departure from equilibrium of the system between 12 and 6 dimensional phase space $\left[1-e^{-(\theta_2-\theta_1)}\right]\approx\theta_2-\theta_1$ multiplied by $Nk$ ($k$ is Boltzmann constant).Now we prove whether $R(t)$) is negative or positive, i.e. that the entropy change from internal interaction is decreasing or increasing, is determined by that internal interaction is negative or positive. Because $f_1(\boldsymbol{x},t)$ and $f_{10}(\boldsymbol{x})$ are probability density and all are greater than zero, and their order of magnitude are the same, so $f_1/f_{10}$ is greater one order of magnitude than $\ln(f_1/f_{10})$. Similarly, because $f_2(\boldsymbol{x},\boldsymbol{x}_1,t)$ is the joint probability density and greater than zero, and satisfied normalization condition $\int f_2(\boldsymbol{x},\boldsymbol{x}_1,t)d\boldsymbol{x}d\boldsymbol{x}_1=1$ and $\int\nabla_p f_2(\boldsymbol{x},\boldsymbol{x}_1,t)d\boldsymbol{p}=0$, hence when $\boldsymbol{p}$ is larger than some critical value, $f_2(\boldsymbol{x},\boldsymbol{x}_1,t)$ will decrease and tend to zero with $\boldsymbol{p}$ increase, so for those $f_2(\boldsymbol{x},\boldsymbol{x}_1,t)$ functions whose part decreasing with $\boldsymbol{p}$ increase is main, $\nabla_p f_2(\boldsymbol{x},\boldsymbol{x}_1,t)$ may be regarded as a minus sign term. Besides, $\int(\nabla_q\phi)\cdot\nabla_p f_2(\boldsymbol{x},\boldsymbol{x}_1,t)d\boldsymbol{p}=0$, and when the system is near the equilibrium, $\nabla_p f_2(\boldsymbol{x},\boldsymbol{x}_1,t)$ is also the same order of magnitude as $\nabla_p f_{20}(\boldsymbol{x},\boldsymbol{x}_1)$, that is $\nabla_p f_2(\boldsymbol{x},\boldsymbol{x}_1,t)\approx\nabla_p f_{20}(\boldsymbol{x},\boldsymbol{x}_1)$. Substituting this result into the penultimate of formula (95) we obtained

$$R(t)\approx Nk\int(\nabla_q\phi)\cdot\left[\frac{f_1(\boldsymbol{x},t)}{f_{10}(\boldsymbol{x})}-\ln\frac{f_1(\boldsymbol{x},t)}{f_{10}(\boldsymbol{x})}\right]\nabla_p f_2(\boldsymbol{x},\boldsymbol{x}_1,t)d\boldsymbol{x}d\boldsymbol{x}_1$$

Under integral of this formula, the second factor $\left[f_1/f_{10}-\ln(f_1/f_{10})\right]$ is plus and the third factor $\nabla_p f_2(\boldsymbol{x},\boldsymbol{x}_1,t)$ is minus, hence whether $R(t)$ is negative or positive is decided by that the first factor $\nabla_q \phi$ is plus or minus. When the internal interaction is repulsive( $\phi > 0, \nabla_q \phi < 0$ ),then $R(t) > 0$ , the entropy increases; conversely, when the internal interaction is attractive( $\phi < 0, \nabla_q \phi > 0$ ),then $R(t) < 0$ ,the entropy decreases. In short, the internal repulsive interaction leads to entropy increase, and the attractive interaction results in entropy decrease.

Now let us consider the physical basis of the entropy change caused by internal interaction in statistical thermodynamic system from expression (96) all-sidedly.

A. The system is nonequilibrium, i.e. $\theta_1 \neq 0$ , $\theta_2 \neq 0$ . Conversely, when the system is equilibrium, i.e. $\theta_1 = 0, \theta_2 = 0$ , then $R = 0$ . It should be pointed out here that entropy decrease $R = 0$ and entropy increase $P$ =0 at equilibrium state are direct conclusions of formulas (96) and (55)(58), it need not any additive condition. Compared with formulas(55)(58), and (96), if expressions (31a)(40a) are applied as the definition of nonequilibrium entropy, although the total form of nonequilibrium entropy evolution equation (44a) does not change, the third term and the fourth term on the right-hand side of equation, i.e. the expressions of the entropy increase rate and the entropy decrease rate of the system change into

$$P'(t) = kD \int f_1(\boldsymbol{x},t) \left[ \nabla_q \ln f_1(\boldsymbol{x},t) \right]^2 d\boldsymbol{x} \geq 0 \tag{58a}$$

$$R'(t) = Nk \int f_2(\boldsymbol{x},\boldsymbol{x}_1,t)(\nabla_q \phi) \cdot \nabla_p \ln f_1(\boldsymbol{x},t) d\boldsymbol{x} d\boldsymbol{x}_1 \tag{96a}$$

Here if we want that $P' = 0$ and $R' = 0$ at equilibrium state, it need $\nabla_q f_1 = 0, \nabla_p f_1 = 0$ . In general, this is not reasonable. This is the reason why expression (31)(40) but not expressions (31a)(40a) are applied as the definition of the nonequilibrium entropy.

B. The departure from equilibrium of the system is inhomogeneous in momentum subspace, i.e. $\nabla_p \theta_1 \neq 0$ . Conversely, when the departure from equilibrium of the system is homogeneous, i.e. $\nabla_p \theta_1 = 0$ , then $R = 0$ .

C. The system is nonlinear and statistically correlated, i.e. $f_2(\boldsymbol{x}, \boldsymbol{x}_1, t) = f_1(\boldsymbol{x}, t) f_1(\boldsymbol{x}_1, t) + g_2(\boldsymbol{x}, \boldsymbol{x}_1, t)$, where $g_2(\boldsymbol{x},\boldsymbol{x}_1,t)$ is two-particle correlation function. Conversely, when the system is nonlinear but statistically independent, i.e. $f_2(\boldsymbol{x}, \boldsymbol{x}_1, t) = f_1(\boldsymbol{x}, t) f_1(\boldsymbol{x}_1, t)$ , then $R = 0$ . Vlasov equation just describes such a system which has neither entropy increase nor entropy decrease[2,3]. If $f_2(\boldsymbol{x},\boldsymbol{x}_1,t) = 0$ , the system is linear, $R$ is also equal to zero.

Here, a relevant fundamental question naturally arise: what is actually the microscopic physical mechanism for the traditional law of entropy increase? According to the result of Boltzmann equation, the entropy increase of the system originates from the short-range repulsive interaction or collision. Making an inference from this result, if the internal interaction is attractive but not repulsive, the entropy of the system should decrease but not increase, then thermodynamic second law described by the entropy increase becomes invalid. However, according to formulas (55)(58) derived by the author, the stochastic diffusion motion of the particle inside the system and inhomogeneous departure from equilibrium of the number density of micro-state are the microscopic physical basis of the macroscopic entropy increase. In other words, the traditional law of entropy increase is caused by the stochastic motion of the particle, but has no direct relation to the internal interaction. Because the stochastic motion is inherent in the microscopic particle of statistical thermodynamic system, the law of entropy increase has universal meaning. Only when there is attractive interaction, the entropy decrease also appears in the system except the law of entropy increase. Both of them coexist in a same system and compete with each other.

**7. Unification of thermodynamic degradation and self-organizing evolution**

The evolution in nature has two directions. One is thermodynamic degradation described by thermodynamic second law, which is degenerative evolution of the system spontaneously tending to the direction increasing the degree of disorder. The other is self-organizing evolution whose type is biological evolution, which is newborn evolution of the system spontaneously tending to the direction increasing the degree of order. How can we unify thermodynamic degradation and self-organizing evolution? This is all along a fundamental problem to be solved in nonequilibrium statistical physics. Now let us give the time rate of change of total entropy in an isolated system and open system, and from these obtained expressions discuss this problem.

At first we present a expression for the time rate of change of total entropy in an isolated system with internal attractive interaction. Integrating both side of nonequilibrium entropy evolution equation (44a) over 6 dimensional phase space and substituting formulas (58) and (96) into this equation, we obtain the expression for the time rate of change of total entropy in an isolated system[18]

$$\frac{\partial_i S}{\partial t} = kD\overline{(\nabla_q \theta_1)^2} + Nk\overline{(\nabla_q \phi)\cdot(\nabla_p \theta_1)\left[1 - e^{-(\theta_2 - \theta_1)}\right]} \tag{97}$$

The first and second terms on the right–hand side of evolution equation (44a) have disappeared from formula (97), because there is no entropy inflow or outflow for an isolated system, and the entropy diffusion only affects the local distribution of entropy density but not affects the total entropy increase or decrease of the system. Formula (97) shows that the time rate of total entropy change of an isolated system (on the left-hand side) is equal to the sum of the formula for the traditional law of entropy increase (the first term on the right-hand side) and the formula for the entropy decrease rate (the second term on the right-hand side). The former is positive and originates from the stochastic motion of microscopic particle inside the system, the latter is negative and comes from the attractive interaction between microscopic particles inside the system. Both of them coexist and countervail each other. Entropy increase destroys the order structure and represents thermodynamic degradation, entropy decrease produces the order structure and expresses self-organizing evolution. It is obvious from formula (97), if there is only the first term of entropy increase, then the entropy of an isolated system only increases and never decreases, i.e. $\partial_i S/\partial t \geq 0$, the system will become more and more disorder with time. However, when the second term of entropy decrease is present and greater than the first term of entropy increase, the time rate of total entropy change of an isolated system is less than zero, i.e. $\partial_i S/\partial t < 0$, hence the self-organizing structure can appears. It can be seen from this that the entropy decrease caused by internal attractive interaction just make an isolated system to be able to overcome the law of entropy increase and produce self-organizing structure. Because the internal interaction in a great many natural systems is neutral, i.e. $\nabla_q \phi = 0$, only the first term on the right-hand side of formula (97) remains. This is another reason that the traditional law of entropy increase is universal. If there is only a repulsive interaction inside the system, i.e. $\nabla_q \phi < 0$, then the second term on the right-hand side of formula (97) is also positive entropy increase. In other words, two forms of entropy increase in both stochastic motion and collision will present simultaneously in this system. Stochastic motion happens in the coordinate space, collision occurs in the momentum space. This system is just described by Boltzmann diffusion equation (15). Here we again see that both stochastic motion and repulsive force result in entropy increase, while attractive force leads to entropy decrease.

Now let us apply formula (97) to discuss one topic of real isolated system.

Suppose two kinds of black and white gas molecules all with same volume and mass mix in an adiabatic container. If there is no interaction force among gas molecules, the only one result after evolution is that two kinds of black and while molecules homogeneously distribute in the whole container. However, when there is attractive interaction force betwwen black molecules or white molecules, considering the role of the entropy decrease on the right-hand side of formula (97), some local region of the container will present higher density of black molecules or white molecules after evolution. Hence an inhomogeneous structure of density distribution produces.

If the system is open, there is entropy inflow or outflow. Integrating both side of nonequilibrium entropy evolution equation (44a) over 6 dimensional phase space and substituting formulas (97) and (96) into this equation, we obtain the expresssion for the time rate of change of total entropy in an open system as follows[18]

$$\frac{\partial_o S}{\partial t} = -\int (\boldsymbol{C} \cdot S_{\text{v}}) \cdot d\boldsymbol{A} + kD\overline{(\nabla_q \theta_1)^2} + Nk\overline{(\nabla_q \phi) \cdot (\nabla_p \theta_1)\left[1 - e^{-(\theta_2 - \theta_1)}\right]} \tag{98}$$

Where $\boldsymbol{C} = \boldsymbol{C}(\boldsymbol{q}, t)$ is average velocity of fluid, $A$ is the surface area of the system, $S_{\text{v}} = \int S_{\text{v}p} d\boldsymbol{p}$ is the entropy density per unit volume. Compared with expression (97), expression (98) is added to a entropy flow term. It shows that the time rate of change of total entropy in an open system (on the left-hand side) is determined together by entropy increase (the second term on the right-hand side), entropy decrease (the third term on the right-hand side) and entropy flow (the first term on the right-hand side). The entropy can flow out to environment from the system owing to the presence of flow term. As a result, the self-organizing structure can emerge from the system only if the sum of entropy outflow and entropy decrease is greater than entropy increase. It can be seen that for self-organizing structure emerging in open system the entropy decrease caused by internal attractive interaction still play promoting role although it no long must be a dominant power. Here we again see that the power of entropy decrease can help open system to resist the low of entropy increase and to produce self-organizing structure. Of course, if the internal interaction is repulsive but not attractive, the third term on the right-hand side of formula (98) is also positive entropy increase. Under such circumstances, if open system want to produce self-organizing structure, it need more entropy flowing out from the system. Hence it is more difficult. If the entropy outflow is so less that the sum of it and entropy decrease is much less than entropy increase, then the total entropy of open system still increase with time.

From expressions (97) (98) it can be seen that on the one hand the entropy in nonequilibrium system always increases (produces), which shows that the power of thermodynamic degradation is eternal; and on the other hand, in addition to the entropy outflow to environment from the system the entropy decrease can appears in nonequilibrium system with internal attractive interaction, which manifest the self-organizing power is accumulating strength for showing off. Both of them coexist in a system and theoretical expression and countervail each other. Hence we can conclude that expressions (97) and (98) have unified thermodynamic degradation for destroying the order structure and self-organizing evolution for producing the order structure.

**8. Approach to equilibrium**

Why nonequilibrium system always approaches to equilibrium? What mechanism is responsible for this process? How to quantitatively describe it? This is also an important problem to be solved in nonequilibrium statistical physics for a long time[3,9]. Because the entropy change represents the evolution direction of nonequilibrium system, now let us solve this problem from the spontaneous change of the entropy in nonequilibrium system. From the existence of entropy density diffusion in nonequilibrium entropy evolution equations (35)(44)(49) and the formula for entropy production rate (55) (58) it is easy to have the idea that the processes of nonequilibrium system approaching to equilibrium is just caused and accomplished by the spontaneous diffusion of the entropy from its high density region to low density region, and the rate for approach to equilibrium is determined by the entropy diffusion rate. The gradient of entropy density decreases gradually and the total entropy of the system increases continuously with the continuous processes of entropy diffusion. At final, the distribution of entropy density in all the system becomes uniform and the entropy diffusion stops, the total entropy changes to maximum, and the system approaches to equilibrium. This mechanism, not only its physical meaning is clear, but also can be used for calculating the relaxation time of actual nonequilibrium system for approach to equilibrium.

Now let us use this idea and method in calculating the entropy density diffusion rate and relaxation time for two actual nonequilibrium systems.

8.1 Free expansion of gas

In previous 5.1.1 section we have gave the entropy production rate for free expansion of gas, now substituting formulas (61)(62) into (41), we obtain the entropy density diffusion rate for free expansion of gas as follows

$$P_d = \frac{\partial^2 S_V}{\partial q^2} = \frac{k}{2t}[(1-\frac{q^2}{2Dt})(1+\ln\frac{f}{f_o})-\frac{q^2}{2Dt}]f \tag{99}$$

It can be seen that the entropy density diffusion rate $P_d$ changes with time $t$ and space $q$.

When the diffusion time

$$t = t_f \geq \frac{l^2}{\pi D} \tag{100}$$

($l$ is the half length of cylinder and very long),the system approaches to equilibrium, entropy diffusion stops, $P_d = 0$ and the entropy production $\Delta_i S$ changes to maximum [ see formula (72)]. Thus $t_f$ is the relaxation time of the system.

8.2 Brownian motion

In previous 5.1.2 section we have gave the entropy production rate for Brownian motion, now substituting formulas (73) (74) into (41), we obtain the entropy density diffusion rate for Brownian motion as follows

$$P_d = D\frac{\partial^2 S_V}{\partial q^2}$$

$$= kDf\left\{-\left[\frac{2\left(q-b(t)\right)^2}{a^2(t)} - \frac{1}{a(t)}\right](1+\ln\frac{f}{f_0}) - \frac{2\left(q-b(t)\right)^2}{a^2(t)} - \frac{1}{a_m} + \frac{4q[q-b(t)]}{a(t)a_m}\right\} \quad (101)$$

This formula shows that the entropy density diffusion rate $P_d$ changes with time $t$ and space $q$.

When the diffusion time

$$t = t_f \geq \beta^{-1} = \frac{a_m}{2D}, \quad (102)$$

the system reaches to equilibrium , the entropy diffusion stops, the entropy density diffusion rate $P_d = 0$, and the entropy production $\Delta_i S$ changes to maximum [seen formula (79)]. Hence $t_f = \beta^{-1}$ is the relaxation time of the system of Brownian motion.

It can be seen from above two topics that the relaxations time $t_f$ of the system approaching to equilibrium is proportional to the square of linear size of the system and inversely proportional to diffusion coefficient.

## 9. Equilibrium ensembles

Equilibrium statistical physics should be a time-independent special part of nonequilibrium statistical physics. However, in the present equilibrium statistical physics, the fundamental probability density is obtained mainly from a fundamental postulate ------the principle of equal probability[2,7,9]. It implies that for an isolated system in a statistical thermodynamic equilibrium state all microscopic states on the same energy surface have an equal probability. Thus following Gibbs, we have the microcanonical distribution[2,7,9]

$$\rho_0 = \rho_0(H) = \Omega^{-1} \quad (103)$$

and the corresponding ensemble is the microcanonical ensemble. Here both the energy $H$ and the number of microstates $\Omega$ are constant.Formula(103) is a equilibrium solution of the Liouville equation, that is

$$\frac{\partial \rho_0}{\partial t} = [H, \rho_0] = 0 \quad (104)$$

But conversely, we cannot solved out formula (103) from equation(104).

Now we still start from the principle of equal probability. When an isolated system is in a statistical thermodynamic equilibrium state, $\rho_0$ is a constant on the same energy surface and independent on space coordinate $\boldsymbol{q}$, hence in the equation (6) (6a) the diffusion term must be zero, i.e. $D\nabla_{\boldsymbol{q}}^2 \rho_0 = 0$. In this case, the Liouville diffusion equation (6) (6a) is reduced to equation (104). This shows that the microcanonical ensemble (103) is also a equilibrium solution of the Liouville diffusion equation (6) (6a). When the microcanonical ensemble is known, we can use the method similar to the present equilibrium statistical physics[2,8,9] to find the canonical and grand canonical ensembles. It can be seen here that equilibrium ensembles are also obtained mainly from postulate and statistics but not dynamics.

**10. Conclusions**

Nonequilibrium statistical physics is considered as a independent main branch of theoretical physics, can we construct a rigorous and unified theory starting from exploring a complete fundamental equation as same as the other main branch of theoretical physics? This problem is related to the direction how to construct the theory. The answer should be positive according to the results of this paper. To propose the stochastic velocity tpye's Langevin equation in 6N dimensional phase space or its equivalent Liouville diffusion equation as the new fundamental equation of nonequilibrium statistical physics, this is only a fundamental assumption. However, it reveals that the law of motion of statistical thermodynamics is expressed by a superposition of both the law of dynamics and stochastic velocity, hence it is essentially different from the law of dynamics. The stochastic diffusion motion of particles is microscopic origin of macroscopic irreversibility.

Starting from this new fundamental equation the BBGKY diffusion equation hierarchy, the Boltzmann collision diffusion equation and the hydrodynamic equations have been derived and presented here. Furthermore, we first wrote down the nonequilibrium entropy evolution equation in 6N , 6 and 3 dimensional phase space, predicted the existence of entropy diffusion . The entropy evolution equation shows that the time rate of change of nonequilibrium entropy density originates together from its drift, diffusion and production in space. From this evolution equation we obtained the formula for entropy production rate (i.e. the law of entropy increase) in 6N and 6 dimensional phase space, the common expression for entropy decrease rate or another entropy increase rate, the expression for unifying thermodynamic degradation and self-organizing evolution, and revealed the entropy diffusion mechanism for approach to equilibrium.

The corresponding quantum Liouville diffusion equation, the quantum entropy evolution equation and the formula for quantum entropy production rate were also presented.

The formula for entropy production rate demonstrates that the macroscopic entropy production of a nonequilibrium physical system is caused by spatially stochastic and inhomogeneous departure from equilibrium of the number density of micro-states. The common expression for entropy decrease rate or another entropy increase rate shows that internal attractive force in noneqilibrium system can result in entropy decrease while internal repulsive force leads to another entropy increase except the traditional entropy increase. The expression for unifying thermodynamic degradation and self-organizing evolution is expressed by the sum of the formula for entropy production rate and the formula for the entropy decrease rate (and the expression for entropy flow), it means that both the traditional entropy increase and the entropy decrease coexist in a system and a theoretical expression and countervail each other. The entropy diffusion mechanism for approach to equilibrium reveals that the processes of an isolated nonequilibrium system approaching to equilibrium is caused and accomplished by the spontaneous diffusion of the entropy from high density region to low density region, and the rate for approach to equilibrium is determined by the entropy diffusion rate.

The microcanonical ensemble is also a equilibrium solution of the Liouville diffusion equation.

All these derivations and results are unified and rigorous from the new fundamental equation without adding any new extra assumption.

Finally, the block diagram for the relation between the new fundamental equation and all the derived and predicted results is shown as follows, where nonequilibrium entropy evolution equation plays a leading role in nonequilibrium entropy theory.

## Block diagram

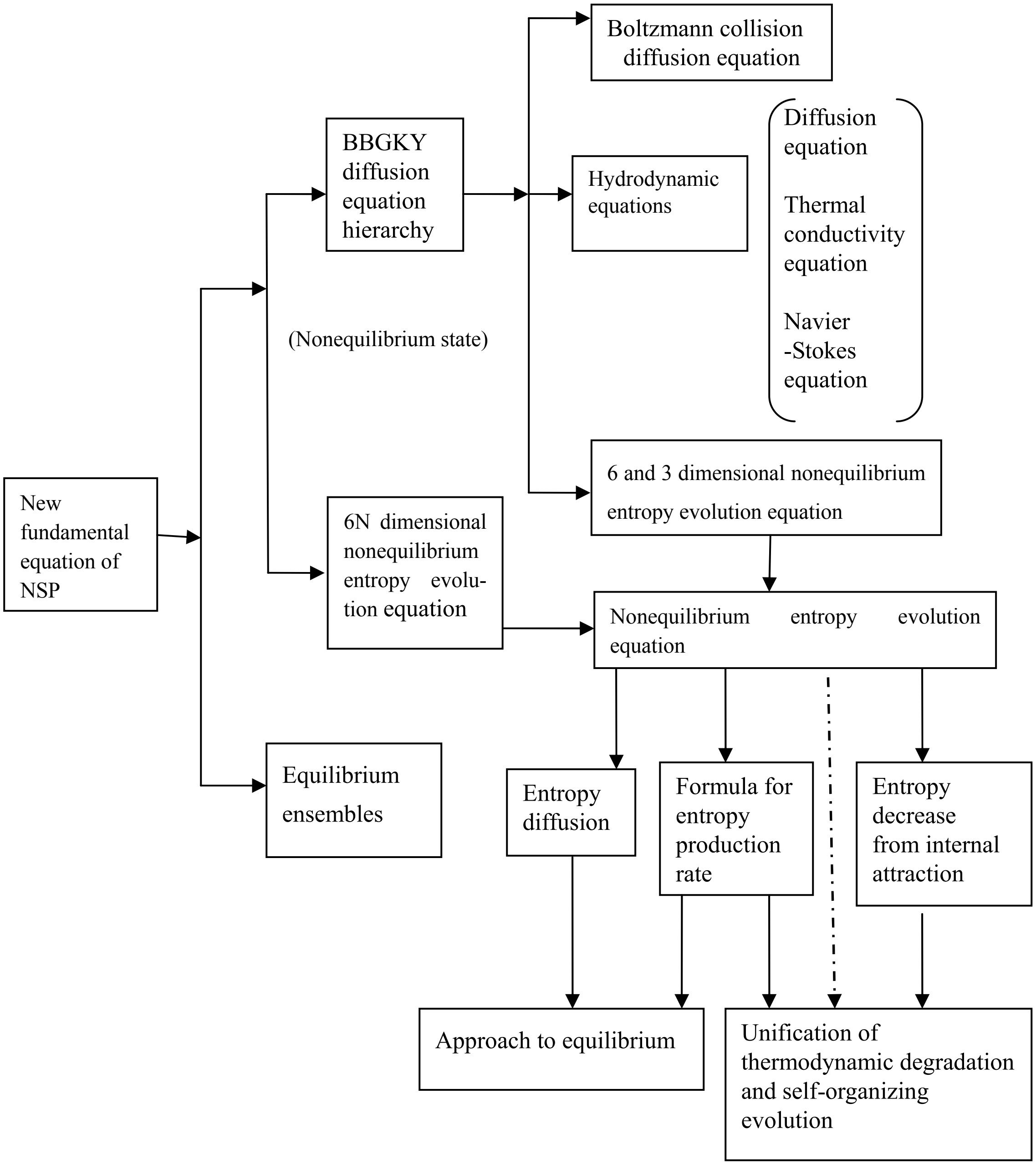